\documentclass[usenatbib]{mn2e}
\pdfoutput=1
\usepackage{graphicx}

\title[Physical properties of \hersch selected galaxies]
{Physical properties of \hersch selected galaxies in a semi-analytic galaxy formation model}

\newcommand{\LCDM}{$\Lambda$CDM }
\newcommand{\msun}{\mbox{${\rm M}_{\odot}$}}
\newcommand{\lsun}{\mbox{${\rm L}_{\odot}$}}

\def\lesssim{\lower.5ex\hbox{$\; \buildrel < \over \sim \;$}}
\def\gtrsim{\lower.5ex\hbox{$\; \buildrel > \over \sim \;$}}
\def\mum{$\mu$m }
\def\mumn{$\mu$m}
\def\redshift{$2 \leq z < 4$}
\newcommand{\hersch}{\textit{Herschel}\ }
\newcommand{\herschn}{\textit{Herschel}}
\newcommand{\spitz}{\textit{Spitzer}\ }

\newcommand{\notice}[1]{{\it {#1}}}

\author[Sami-Matias Niemi et al.]
{
Sami-Matias Niemi$^{1,2}$\thanks{E-mail: sammy@sammyniemi.com (SMN)}, 
Rachel S. Somerville$^{1,3}$,
Henry C. Ferguson$^{1}$,	
\newauthor{Kuang-Han Huang$^{3}$,
Jennifer Lotz$^{1}$ and
Anton M. Koekemoer$^{1}$}
\\
$^{1}$STScI, 3700 San Martin Drive, Baltimore, MD 21218, United States\\
$^{2}$University of Turku, Department of Physics and Astronomy, Tuorla
Observatory, V\"ais\"al\"antie 20, Piikki\"o, Finland\\
$^{3}$Department of Physics and Astronomy, Johns Hopkins University, Baltimore, MD 21218, United States\\
}

\begin{document}

\date{Released 2011}

\pagerange{\pageref{firstpage}--\pageref{lastpage}} \pubyear{2011}

\maketitle

\label{firstpage}

\begin{abstract}
We make use of a semi-analytic cosmological model that includes simple
prescriptions for dust attenuation and emission to make predictions for the
observable and physical properties of galaxies that may be detected by the
recently launched \hersch Space Observatory in deep fields such as
GOODS-\herschn. We compare our predictions for differential galaxy number counts
in the \hersch PACS (100 and 160 \mumn) and SPIRE (250, 350, and 500 \mumn)
bands with available observations. We find very good agreement with the counts
in the PACS bands, for the overall counts and for galaxies binned by redshift at
$z<2$. At $z > 2$ our model underpredicts the number of bright galaxies in the
PACS bands by a factor of ten. The agreement is much worse for all three SPIRE
bands, and becomes progressively worse with increasing wavelength. We discuss a
number of possible reasons for these discrepancies, and hypothesize that
the effect of blending on the observational flux estimates is likely to be the
dominant issue. We note that the PACS number counts are relatively robust to
changes in the dust emission templates, at least for the three sets of templates
that we have tested, while the predicted SPIRE number counts are more template
dependent especially at low redshift. We present quantitative predictions for
the relationship between the observed PACS 160 and SPIRE 250 \mum fluxes and
physical quantities such as halo mass, stellar mass, cold gas mass, star
formation rate, and total infrared (IR) luminosity, at different redshifts. We
also present predictions for the radial sizes of \herschn-selected disks at high
redshift $( z > 2)$ and find reasonable agreement with the available
observations. Finally, we present quantitative predictions for the correlation
between PACS 160 \mum flux and the probability that a galaxy has experienced a
recent major or minor merger. Although our models predict a strong correlation
between these quantities, such that more IR-luminous galaxies are more likely to
be merger-driven, we find that a significant fraction (more than half) of all
high redshift IR-luminous galaxies detected by \hersch are able to attain their
high star formation rates without enhancement by a merger.
\end{abstract}

\begin{keywords}
methods: numerical - galaxies: formation - galaxies: evolution -
galaxies: mergers - galaxies: star formation - galaxies: infrared
selected
\end{keywords}

\section{Introduction}\label{s:intro}

One of the most important discoveries from extragalactic observations in the
mid- and far-infrared has been the identification of luminous and ultra-luminous
infrared bright galaxies (LIRGs; $L_{\mathrm{IR}} > 10^{11}L_{\odot}$ and
ULIRGs; $L_{\mathrm{IR}} > 10^{12}L_{\odot}$, respectively). These objects have
been studied extensively in the literature \citep[e.g.][]{Rieke:1972p1101,
Harwit:1987p1111, Sanders:1988p1131, Sanders:1991p1151, Condon:1991p1127,
Auriere:1996p1135, Duc:1997p1147, Genzel:1998p1155, Lutz:1998p1152,
Rigopoulou:1999p1114, RowanRobinson:2000p1143, Genzel:2001p1106,
Colina:2001p1123, Colbert:2006p1150, Dasyra:2006p1119,
HernanCaballero:2009p1137, Magdis:2011p1146}, and found to emit more
energy at infrared wavelengths ($\sim 5 - 500$ \mumn) than at all other
wavelengths combined \citep[e.g.][]{Sanders:1996p1085}. Even though the luminous
infrared galaxies are rare objects in the local Universe
\citep[e.g.][]{Lagache:2005iw}, reasonable assumptions about the lifetime of the
infrared phase \citep[see, e.g.][and references therein]{Farrah:2003p1149}
suggest that a substantial fraction of all massive galaxies pass through a stage
of intense infrared emission during their lifetime \citep[][and references
therein]{Sanders:1996p1085}. Consequently, the majority of the most luminous
galaxies in the Universe emit the bulk of their energy in the far-infrared.
Furthermore, LIRGs and ULIRGs dominate the cosmic star formation rate density at
$z \sim 1 - 2$, accounting for 70 per cent of the star formation activity at
these epochs \citep[e.g.][]{LeFloch:2005p1121}. This makes the IR an extremely
interesting wavelength regime to study, especially in the context of the cosmic
star formation history, and galaxy formation and evolution.

Understanding IR bright galaxies is especially important because light from
bright, young blue stars is often attenuated by dust \citep[for a review
see][]{Calzetti:2001hr}. The observed rest-frame ultra-violet (UV) light of a
galaxy may therefore provide a biased view of the star formation rate in the
galaxy. For example, the global star formation rate required to explain the
far-infrared and submillimetre background appears to be higher than that
inferred from the data in the UV-optical \citep[e.g.][and references
therein]{1998Natur.394..241H}. The dust attenuated light from blue stars is
however not lost, but is re-radiated at IR wavelengths. As a result, a large
fraction of the radiation from star formation is radiated not at UV or optical
but at IR rest-frame wavelengths. 

The European Space Agency's \hersch Space Observatory \citep{Pilbratt:2010p1057}
was launched in May of 2009. It observes the Universe at infrared wavelengths,
from 60 to 670 \mumn, and opens up a huge region of new parameter space for
surveys in area, depth and wavelength. Consequently, a primary goal of \hersch
is to explore the evolution of obscured galaxies. With \herschn, we can finally
probe the IR light of high-redshift galaxies from 70 to 160 and 250 to 500 \mum
with the Photodetector Array Camera \& Spectrometer
\citep[PACS;][]{Poglitsch:2010p1058} and Spectral and Photometric Imaging
Receiver \citep[SPIRE;][]{Griffin:2010p1059}, respectively. Unfortunately, at
such long wavelengths, contamination and crowding, rather than the depth of the
observations, often becomes a limiting factor
\citep[e.g.][]{1998MNRAS.296L..29B, Brisbin:2010ep, Lacey:2010hp}. As a result,
the interpretation of IR observations can be less than straightforward. This is
especially true when probing galaxies at high redshift. Despite the potential
complications, several deep \hersch observations have been performed
\citep[e.g.][]{Oliver:2010fp} since the launch of the observatory and more will
likely follow.

Early \hersch observations and studies have already generated some very
interesting results, such as the conclusions that the majority of the detected
high-redshift galaxies are large and massive spiral galaxies
\citep[e.g.][]{Cava:2010p1050} with extremely high star formation rates
\citep[e.g.][]{Brisbin:2010ep}. Furthermore, the minimum dark matter halo mass
of IR bright galaxies was recently constrained to be about $3 \times
10^{11}$\msun \citep{Amblard:2011p1070}, while the average dust temperature of
H-ATLAS sources was found to be $28 \pm 8$ K \citep{Amblard:2010fn}. However,
because of the very large beam of the \hersch telescope and the difficulty of
making unique associations between optically detected galaxies and the PACS or
SPIRE sources, many open questions about the physical nature of the populations
detected by \hersch remain: for example, what are their stellar masses, sizes,
and morphologies (spheroid or disk dominated)? Are they dynamically relaxed
galaxies, or interacting or merging systems?

One can turn to theoretical modelling to aid in the interpretation of these
observations, and to make predictions about the nature of the objects that
\hersch should detect at various redshifts. In the IR, all the theoretical
models in the literature can be divided into two broad classes:
\textit{phenomenological} models and \textit{cosmological} models. The
phenomenological models (sometimes called ``backwards evolution'' models) are
based on observed luminosity functions, and assume simple functional forms to
describe the evolution of galaxy luminosity functions or SFR with redshift.
Template spectral energy densities (SEDs) based on observed galaxies are used to
transform between different wavelengths \citep[e.g.][]{Dale:2001p1087,
Dale:2002bo, Lagache:2003p1139}. In contrast, in semi-analytic cosmological
models, the evolution of galaxy properties is predicted based on the framework
of the hierarchical structure formation, or Cold Dark Matter (CDM) theory. In
these simulations, the evolution of structure in the dark matter component is
characterized via ``merger trees'', and simplified prescriptions are used to
model the main physical processes such as cooling and accretion of gas, star
formation, chemical evolution, stellar, supernovae and active galactic nuclei
feedback \citep[e.g.][]{Croton:2006ew, Somerville:2008p759}. Based on the
resulting predicted star formation and chemical enrichment histories, the SED
for unattenuated light from stars (which dominates the SED shortwards of about 3
$\mu$m) may be calculated using stellar population synthesis models
\citep[e.g.][and references
therein]{1976ApJ...203...52T,Worthey:1994fh,1995ApJS...96....9L,
Bruzual:2003ck,2005MNRAS.362..799M, 2009ApJ...699..486C}. 

Modelling the impact of dust attenuation, and computing the SED of the dust
emission, are less straightforward. A standard assumption is that all energy
absorbed by dust is re-radiated. One can then break the problem into two parts:
computing the amount of light absorbed (and scattered) by dust as a function of
wavelength, and computing the amount of re-radiated light emitted by dust as a
function of wavelength. One approach is to couple the semi-analytic models with
a radiative transfer (RT) code \citep{Granato:2000p1073, Baugh:2005dp,
2007MNRAS.382..903F, Lacey:2008p1107, Lacey:2010hp} to compute the dust
attenuation and scattering, and a detailed dust model, that assumes a specific
dust composition and set of grain properties, to compute the dust emission SEDs.
Another approach is to use analytic models relating the attenuation to the
column density of dust (assumed to be traced by metals in the cold gas) in
galactic disks \citep[e.g.][]{guiderdoni:87}. To compute the dust emission SED,
one can then use empirical dust SED templates based on analytic dust models
calibrated to observations, or observed galaxies
\citep[][]{Guiderdoni:1998p1140,devriendt:99,devriendt:00}.

The former approach (full RT+dust model) has the clear advantage of providing
more detailed and accurate predictions of the galaxy SEDs, given the input
assumptions, and certainly provides a better estimate of galaxy-to-galaxy
scatter in the SED properties. It has the disadvantage of being computationally
quite expensive, and perhaps not merited given that semi-analytic models do not
provide detailed information about the relative geometry of stars and dust in
galaxies, nor about the composition or properties of the dust, which may vary
with cosmic time or environment. \citet{fontanot:09a} and \citet{fontanot:11}
have shown that the analytic recipes for predicting dust attenuation and the
empirical template approach for dust emission provide reasonably good agreement,
in most cases, with statistical quantities such as galaxy luminosity functions
and counts as computed with the full RT approach using the GRASIL code
\citep{silva:98}. Moreover, this analytic approach to dust attenuation and
emission was adopted by \citet[][hereafter S11]{somerville:11} and
\citet{gilmore:11}, who showed that it provided good agreement with observed
galaxy luminosity functions for rest-UV to NIR wavelengths and redshifts $0 < z
< 4$. This approach also produced fairly good agreement with observational
estimates of the {\em total} IR luminosity function at $0 < z < 2$, and with the
observed Extragalactic Background Light. However, these models produced poorer
agreement at longer wavelengths ($\lambda \gtrsim 8 \mu$m), particularly for
luminous higher redshift ($z\gtrsim 2$) galaxies. Thus, the limitations of our
approach should be kept in mind. Part of the goal of this work is to determine
where improvements to this kind of analytic approach must be made by confronting
the models with recent observations from Herschel.

In this work we use the SAM introduced in \cite{Somerville:1999p762}, with
significant updates as presented by \citet{Somerville:2008p759},
\citet{2009MNRAS.397..802H}, and S11, and undertake a study of galaxies that
should be detected by current \hersch surveys both locally and at high redshift.
We focus on the populations that can be detected in the relatively deep surveys
being carried out with \herschn, in particular the GOODS-\hersch project (PI D.
Elbaz). The redshift range \redshift \ is of special interest because 1) it
spans the peak of the cosmic star formation activity in the Universe 2) it
probes what are probably the earliest epochs for which individual galaxies can
be detected with \hersch and 3) many open questions remain about the nature of
galaxy populations, especially luminous IR galaxies, at these epochs.

This paper is organised as follows. In Section \ref{s:model}, we provide a brief
overview of the semi-analytic galaxy formation model used in our study, the
model for dust attenuation and emission, how we construct our mock light cone,
and our sample selection criteria. In Section \ref{s:observable_properties} we
present our predictions for observable properties of \hersch galaxies such as
counts and luminosity functions. In Section \ref{s:physical_properties}, we show
predictions for the physical properties of \herschn-selected galaxies at
different redshifts. Finally we summarise our results and conclude in Section
\ref{s:conclusions}.

\section{The Semi-Analytic Model}\label{s:model}

\subsection{The Semi-analytic Galaxy Formation Model}\label{ss:simulations}

We compute the formation and evolution of galaxies within the $\Lambda$CDM
cosmology using the semi-analytic galaxy formation model (SAM) introduced in
\cite{Somerville:1999p762} and described in detail in
\cite{Somerville:2008p759}, hereafter S08. Our model also includes the modified
recipe for bulge formation and starburst efficiency described in
\citet{hopkins:09}. In summary, the S08 SAM includes the following physically
motivated recipes: 1) growth of structure in the dark matter component in a
hierarchical clustering framework, as characterized by ``merger trees''; 2) the
shock heating and radiative cooling of gas; 3) conversion of cold gas into stars
according to an empirical ``Kennicutt-Schmidt'' relation; 4) the evolution of
stellar populations; 5) feedback and metal enrichment of the Interstellar (ISM)
and Intracluster Medium (ICM) from supernovae explosions; 6) two modes
(``quasar'' and ``radio'' mode) of black hole growth and feedback from active
galactic nuclei; 7) starbursts and morphological transformation by galaxy
mergers. The SAM can be used to predict physical quantities such as stellar and cold gas
masses and metallicities of galaxies, their star formation and merger histories,
and their luminosities and sizes.

The merging histories (or merger trees) of dark matter haloes are constructed
based on the Extended Press-Schechter formalism using the method described in
\cite{Somerville:1999p761}, with improvements described in S08. These merger
trees record the growth of dark matter haloes via merging and accretion, with
each ``branch'' representing a merger of two or more haloes. In this work we
follow each branch back in time to a minimum progenitor mass of $10^{9}$\msun.
This mass scale is sufficient to resolve the formation histories of the
relatively bright, massive galaxies that we study in this work, which occupy
halos with masses greater than a few $\times 10^{11}\msun$.

In the SAM of S08 the star formation occurs in two modes, a ``quiescent'' mode
in isolated disks, and a merger-driven ``starburst'' mode. Star formation in
isolated disks is modelled using the empirical Kennicutt-Schmidt relation
\citep{Kennicutt:1989p976}, assuming that only gas above a fixed critical
surface density is eligible to form stars. The efficiency and timescale of the
merger driven burst mode is a function of merger ratio and the gas fractions of
the progenitors, and is based on the results of hydrodynamic simulations
\citep{2006ApJ...645..986R, 2009MNRAS.397..802H}. 

In this work we have assumed a standard \LCDM universe and a Chabrier stellar
initial mass function \citep[IMF;][]{2003PASP..115..763C}. We adopt the
following cosmological parameter values: $\Omega_m = 0.28$, $\Omega_{\Lambda} =
0.72$, $H_0 = 70.0$, $\sigma_8 = 0.81$ and $n_{s} = 0.96$, which are consistent
with the five year Wilkinson Microwave Anisotropy Probe results
\citep{2009ApJS..180..330K} and was also adopted by S11. The adopted baryon
fraction is $0.1658$.

In the following we summarise the main aspects of the modelling of the dust
attenuation and emission. A more detailed account can be found in S11.

\subsubsection{Model for Dust Attenuation}

Our model for dust extinction is based on the approach proposed by
\citet{guiderdoni:87}, combined with the model proposed by
\cite{Charlot:2000p1068}. As in the Charlot \& Fall model, we consider
extinction by two components, one due to the diffuse dust in the disc and
another associated with the dense `birth clouds' surrounding young star forming
regions. The $V$-band, face-on extinction optical depth of the diffuse dust is
given by
\begin{equation}
\tau_{V,0} = \frac{\tau_{\mathrm{dust,0}}\, Z_{\mathrm{cold}}\, m_{\mathrm{cold}}}{r_{\mathrm{gas}}^{2}} \quad ,
\end{equation}
where $\tau_{\mathrm{dust,0}}$ is a free parameter, $Z_{\mathrm{cold}}$ is the
metallicity of the cold gas, $m_{\mathrm{cold}}$ is the mass of the cold gas in
the disc, and $r_{\mathrm{gas}}$ is the radius of the cold gas disc, which is
assumed to be a fixed multiple of the stellar scale length.

To compute the actual extinction we assign a random inclination to each galaxy
and use a standard `slab' model; i.e. the extinction in the $V$-band for a
galaxy with inclination $i$ is given by:
\begin{equation}
A_V = -2.5 \log_{10} \left [ \frac{1 - \exp \left ( \frac{-\tau_{V,0}}{\cos(i)} \right )}{\frac{\tau_{V,0}}{\cos(i)}} \right ] \quad .
\end{equation} 
Additionally, stars younger than $t_{\rm BC}$ are enshrouded in a cloud of dust
with optical depth $\tau_{\mathrm{BC,V}}=\mu_{\mathrm{BC}}\, \tau_{V,0}$, where
we treat $t_{\rm BC}$ and $\mu_{\mathrm{BC}}$ as free parameters.  Finally, to
extend the extinction estimate to other wavebands, we assume a starburst
attenuation curve \citep{1997AJ....113..162C, Calzetti:2001hr} for the diffuse
dust component and a power-law extinction curve
$A_{\lambda}\propto(\lambda/5500$\AA$)^n$, with $n=0.7$, for the birth clouds
\citep{Charlot:2000p1068}. Our results are insensitive to minor modifications of
the underlying extinction curve for the cirrus extinction, e.g. adoption of a
Galactic or SMC-type extinction curve instead of Calzetti.

\subsubsection{Dust Parameters}

There are three free parameters that control the dust attenuation in our model:
the normalization of the face-on $V$-band optical depth
$\tau_{\mathrm{dust,0}}$, the opacity of the birth clouds relative to the cirrus
component $\mu_{\mathrm{BC}}$, and the time that newly born stars spend
enshrouded in their birth clouds, $t_{\rm{BC}}$. We first set
$\tau_{\mathrm{dust,0}}$ by matching the normalization of the observed
relationship between $L_{\rm{dust}}/L_{\rm UV}$ vs. bolometric luminosity
$L_{\rm bol}$ for nearby galaxies, where $L_{\rm dust}$ is the total luminosity
absorbed by dust and re-emitted in the mid- to far-IR and $L_{\rm UV}$ is the
luminosity in the far-UV ($\sim 1500$ \AA). Using a value of
$\tau_{\mathrm{dust,0}}=0.2$, S11 found good agreement with these observations,
and also with the observed optical through NIR luminosity functions in the local
Universe. We also adopt this same value, $\tau_{\mathrm{dust,0}}=0.2$.

The birth cloud parameters $\mu_{\mathrm{BC}}$ and $t_{\rm BC}$ mainly control
the attenuation of UV light relative to longer wavelengths. S11 show that in the
local Universe $(z = 0)$, the $g$ through $K$-band luminosity functions are
insensitive to the birth cloud parameters, while the FUV through $u$-bands are
quite dependent on them. S11 adjusted the parameters to match the $z=0$ FUV and
NUV observed luminosity functions, finding good agreement with
$\mu_{\mathrm{BC}}=4.9$ and $t_{\rm BC}=2\times 10^{7}$ yr, which we also adopt
in our study. However, S11 found that it was not possible to reproduce the
observed rest-UV and optical luminosity functions at high redshift with fixed
values of these parameters. Other studies \citep[e.g.][]{lofaro:09, guo-white:09}
have reached similar conclusions. Moreover, there is direct observational
evidence that galaxies are less extinguished at high redshift for a given
bolometric luminosity \citep{reddy:10}. Following S11, we therefore adopt a
simple redshift dependence in all three dust parameters, which is adjusted in
order to reproduce the observed rest-frame UV and optical luminosity functions
out to $z\sim5$. Our adopted scalings are: $\tau_{\mathrm{dust,0}}(z) =
\tau_{\mathrm{dust,0}}(1+z)^{-1}$, and both $\mu_{\mathrm{BC}}$ and $t_{\rm BC}$
scale with $z^{-1}$ above $z = 1$.

\subsubsection{Model for the Dust Emission}

Using the formalism presented above we can compute the total fraction of the
energy emitted by stars that is absorbed by dust, over all wavelengths, for each
galaxy in our simulation. We then assume that all of this absorbed energy is
re-radiated in the IR (we neglect scattering), and thereby compute the total IR
luminosity of each galaxy $L_{\rm IR}$. We then make use of dust emission
templates to determine the SED of the dust emission, based on the hypothesis
that the shape of the dust SED is well-correlated with $L_{\rm IR}$. The
underlying physical notion is that the distribution of dust temperatures is set
by the intensity of the local radiation field; thus more luminous or actively
star forming galaxies should have a larger proportion of warm dust, as
observations \citep[e.g.][]{Sanders:1996p1085} seem to imply.

There are two basic kinds of approaches for constructing these sorts of
templates. The first is to use a dust model along with either numerical or
analytic solutions to the standard radiative transfer equations to create a
library of templates, calibrated by comparison with local prototypes. This
approach was pioneered by \cite{Desert:1990p1069}, who posited three main
sources of dust emission: polycyclic aromatic hydrocarbons (PAHs), very small
grains and big grains. In this approach, the detailed size distributions are modelled using free
parameters, which are calibrated by requiring the model to fit a set of
observational constraints, such as the extinction or attenuation curves,
observed IR colours and the IR spectra of local galaxies.

The second approach is to make direct use of observed SEDs for a set of
prototype galaxies and to attempt to interpolate between them
\citep[e.g.][]{Chary:2001jf,Dale:2002bo,lagache:04}. In this work we make use of
the empirical SED templates recently published by \cite{Rieke:2009p1088},
hereafter R09. They constructed detailed SEDs from published ISO, IRAS and
NICMOS data as well as previously unpublished IRAC, MIPS and IRS observations.
They modelled the far infrared SEDs assuming a single blackbody with
wavelength-dependent emissivity. The R09 library includes fourteen SEDs covering
the $5.6 \times 10^9 L_\odot<L_{\mathrm{IR}}< 10^{13} L_\odot$ range. To explore
the sensitivity of our results to the details of the template set, we also
present results for the \citet[][]{Chary:2001jf} templates, hereafter CE01, and
for the templates recently presented by \citet[][]{chary-pope:10}, hereafter
CP11. The latter are calibrated to reproduce the SEDs of high redshift galaxies,
which appear to deviate somewhat from the local templates (for a detailed
discussion, see CP11).

\subsection{The Mock Light Cone}\label{ss:mock_light_cone}

We use the SAM described in Section \ref{ss:simulations} to generate a mock
light cone of simulated galaxies. Our simulated light cone is $100$ times the
size of the Great Observatories Origins Deep Survey (GOODS) on the northern sky,
which covers $\sim 160$ arcminutes squared in the sky. We chose to simulate a
light cone that covers $100$ times larger area than the GOODS-N to improve the
statistics, as luminous IR galaxies are relatively rare. We further limit our
light cone of simulated galaxies to range from zero to six in redshift space.
This limitation is arbitrary, however, we will show below that the apparent IR
flux of simulated galaxies drops steeply as a function of redshift, so that it
is unlikely that any galaxies beyond this redshift could be detected by
\herschn.

Our simulated light cone contains in total $23,980,599$ galaxies. The dark
matter halo masses range from $\sim 2.7 \times 10^{9}$ to $\sim 4.5 \times
10^{14}$\msun. Consequently, our lightcone contains a variety of dark matter
haloes from light subhaloes to large and massive cluster sized haloes. As a
result, these simulated dark matter haloes host a variety of different types of
galaxies from small dwarf galaxies to giant ellipticals. The stellar masses of
our simulated galaxies range roughly five orders in magnitude from $10^{7}$ to
$\sim 4.6 \times 10^{12}$\msun.

We use the full mock light cone from redshift zero to six when comparing the
predicted number counts to observational constraints (Section
\ref{ss:number_counts}) and for predictions of the galaxy luminosity functions
(Section \ref{ss:luminosity_function}). We also use all galaxies in our full
mock light cone when making predictions for physical properties as a
function of redshift. However, in some cases we select a subsample that contains
only high-redshift IR bright galaxies.

\subsubsection{Sample of High-Redshift luminous IR Galaxies}\label{ss:sample}

\begin{figure*}
\begin{minipage}{18cm}
\begin{center}
\includegraphics[width=\columnwidth]{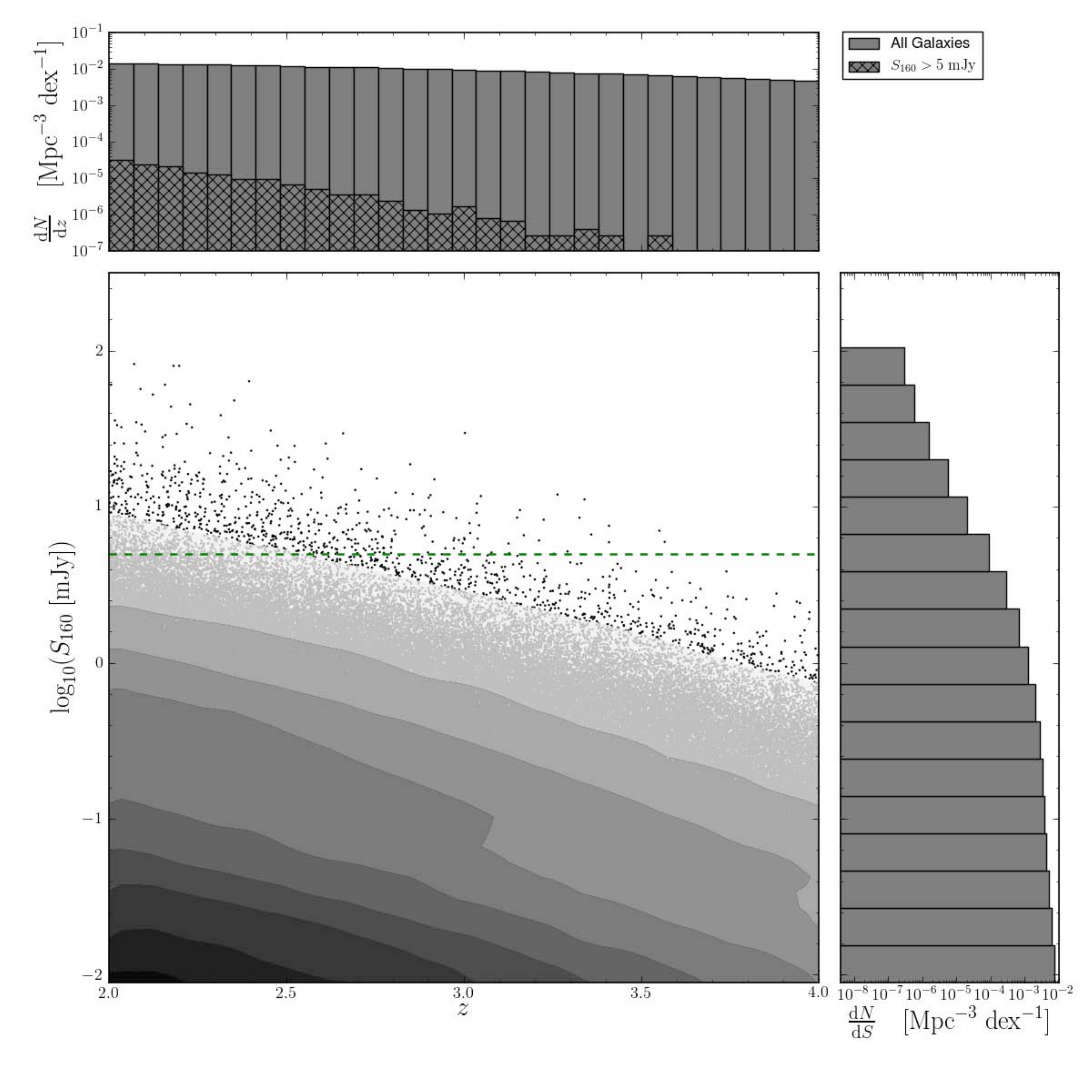}
\caption{PACS 160 \mum flux, $S_{160}$, in mJy as a function of redshift $z$ for
all simulated galaxies. The histograms show projections of the scatter plot. The
solid bars in the upper histogram show co-moving number density
$(\frac{\mathrm{d}N}{\mathrm{d}z})$ of all galaxies, while the hatched bars show
that of simulated galaxies with $S_{160} > 5$ mJy. The right-hand side histogram
shows co-moving number density $(\frac{\mathrm{d}N}{\mathrm{d}S})$ as a function
of PACS 160 \mum flux. The horizontal green dashed line is at $5$ mJy, which
corresponds roughly to the detection limit in PACS 160 \mum observations in the
GOODS-N.}
\label{fig:pacs_selection}
\end{center}
\end{minipage}
\end{figure*}

For part of our analysis, we focus on a sample with criteria that are motivated
by the desire to compare with high redshift galaxies in the GOODS-\hersch
survey. For this sample, we select all simulated galaxies from our light cone
that are in the redshift range \redshift\ and above the detection limit of
either PACS at the $160$ \mum or SPIRE at the $250$ \mum bands in the GOODS
North. For simplicity, we assume the limits for both instruments to be $\sim 5$
mJy \citep[see, e.g.][]{Elbaz:2010do}, although we note that the actual
detection limit in the PACS band may be somewhat lower ($\sim 3.5 - 4.5$ mJy).
These two bands were chosen to represent the two different instruments and to be
close to the peak of the thermal IR emission in our chosen redshift range. 
No other selection criteria, such as stellar mass or optical flux cut, were
applied, except the ones enforced by the mass resolution.

Selected in this way, our high-redshift sample comprises in total 1154 $(3948)$
galaxies when the selection is done using the PACS 160 (SPIRE 250) \mum band.
Note however that our simulated light cone is $100$ times larger than the
GOODS-N --- hence one may infer that we would expect $\sim 10$ PACS 160 \mum and
$\sim 40$ SPIRE 250 \mum detected galaxies in a field with the size and depth of
GOODS-N. However, cosmic variance is probably large for these objects, and
therefore this number is uncertain. We return to this matter in the following
Sections.

The PACS 160 \mum selected high-redshift sample consists of luminous IR galaxies
with bolometric IR luminosity ranging from $\sim 1 \times 10^{12}$ to $\sim 2
\times 10^{13}$\lsun. The average bolometric luminosity $\bar{L}_{\textrm{IR}}
\sim 2.46 \times 10^{12}$\lsun. The SPIRE 250 \mum selected high-redshift sample
consists of IR bright galaxies with the bolometric IR luminosity ranging from
$\sim 5\times 10^{11}$ to $\sim 2 \times 10^{13}$\lsun, while the average
bolometric IR luminosity $\bar{L}_{\textrm{IR}} \sim 1.75 \times 10^{12}$\lsun.
We can therefore readily conclude that all of the galaxies in the PACS selected
sample are ULIRGs, while in the SPIRE selected sample they are at least LIRGs,
while most, i.e. more than $90$ per cent, can be defined as ULIRGs. The simple
single PACS 160 \mum or SPIRE 250 \mum band selection is a good indicator for
the total IR luminosity because in our selected redshift range we are always
near the peak of the IR flux. Consequently, our sample of high-redshift galaxies
is well suited to study the physical properties of high redshift (U)LIRGs that
are potentially observable in fields such as the GOODS North and South.

Figure \ref{fig:pacs_selection} shows the PACS 160 \mum flux, $S_{160}$, in mJy
as a function of redshift $z$ for all simulated galaxies as well as the
co-moving number density of simulated galaxies as a function of PACS 160 \mum
flux and redshift. The horizontal green dashed line in the scatter plot is drawn
at $5$ mJy, which is roughly the same as the detection limit in the GOODS-N. 
Note, however, that the confusion limit in the GOODS-N and in comparable fields
can be higher than the detection limit, especially at longer wavelengths. For
example, a confusion limit $\sim 19.1 \pm 0.6$ mJy at the 250 \mum SPIRE band
has been quoted in literature \citep[][]{2010A&A...518L...5N}. Obviously,
observations that are of the same depth, but on less crowded fields could reach
down to the mJy level. Moreover, observations in gravitational lensed fields can
reach well below the derived confusion limit, as has already been shown by
\cite{Altieri:2010p1044}, thus providing a motivation for our sample selection
and the usage of $5$ mJy. We note that the Fig. \ref{fig:pacs_selection} would
be quantitively similar if we used the SPIRE 250 \mum band instead of the PACS
160 \mum flux, although the SPIRE-selected sample extends to slightly higher
redshift (a few galaxies are predicted above 5 mJy up to $z \sim 4.0$).

Figure \ref{fig:pacs_selection} shows that the number density of luminous IR
galaxies drops quickly above redshift three, hence, observations comparable to
those in GOODS-N are unlikely to detect galaxies above redshift four.  However,
as noted above, these galaxies are likely to be strongly clustered, with a large
field-to-field variance, implying that the maximum redshift may also vary from
field to field.  Unfortunately, due to the fact that we adopt the Extended
Press-Schechter formalism to generate our merger trees, we do not have
information on the spatial locations of our galaxies and therefore we cannot
properly quantify the effects of cosmic variance. We defer this to a future
study, where we plan to use lightcones extracted from $N$-body simulations and
will therefore have information on the clustering properties of our galaxies.

\section{Observable Properties} \label{s:observable_properties}

\subsection{Number Counts and Redshift Distributions}\label{ss:number_counts}

The most basic statistic describing a galaxy population is the number counts,
i.e. the number density on the sky of galaxies as a function of observed flux.
The number counts at far-infrared and sub-mm wavelengths are well known to
exhibit strong evolution \citep[][and references therein]{Oliver:2010fp}.
Unfortunately, the \hersch beam is broad compared to the number density of
sources, and thus the maps are often confused. This confusion means that care
has to be taken when estimating number counts from \hersch data: observations
must in general be corrected for flux boosting and incompleteness \citep[see,
e.g.,][and references therein]{Clements:2010ce, Oliver:2010fp}. For example, the
faint flux densities may be overestimated due to the classical flux boosting
effect \citep[e.g.][]{2010A&A...516A..43B}. In the following sections we
therefore compare our model to existing observations and make predictions for
higher redshifts and lower flux levels than what the current observations have
probed.

\subsubsection{Number Counts in the PACS and SPIRE Bands}

\begin{figure}
\includegraphics[width=\columnwidth]{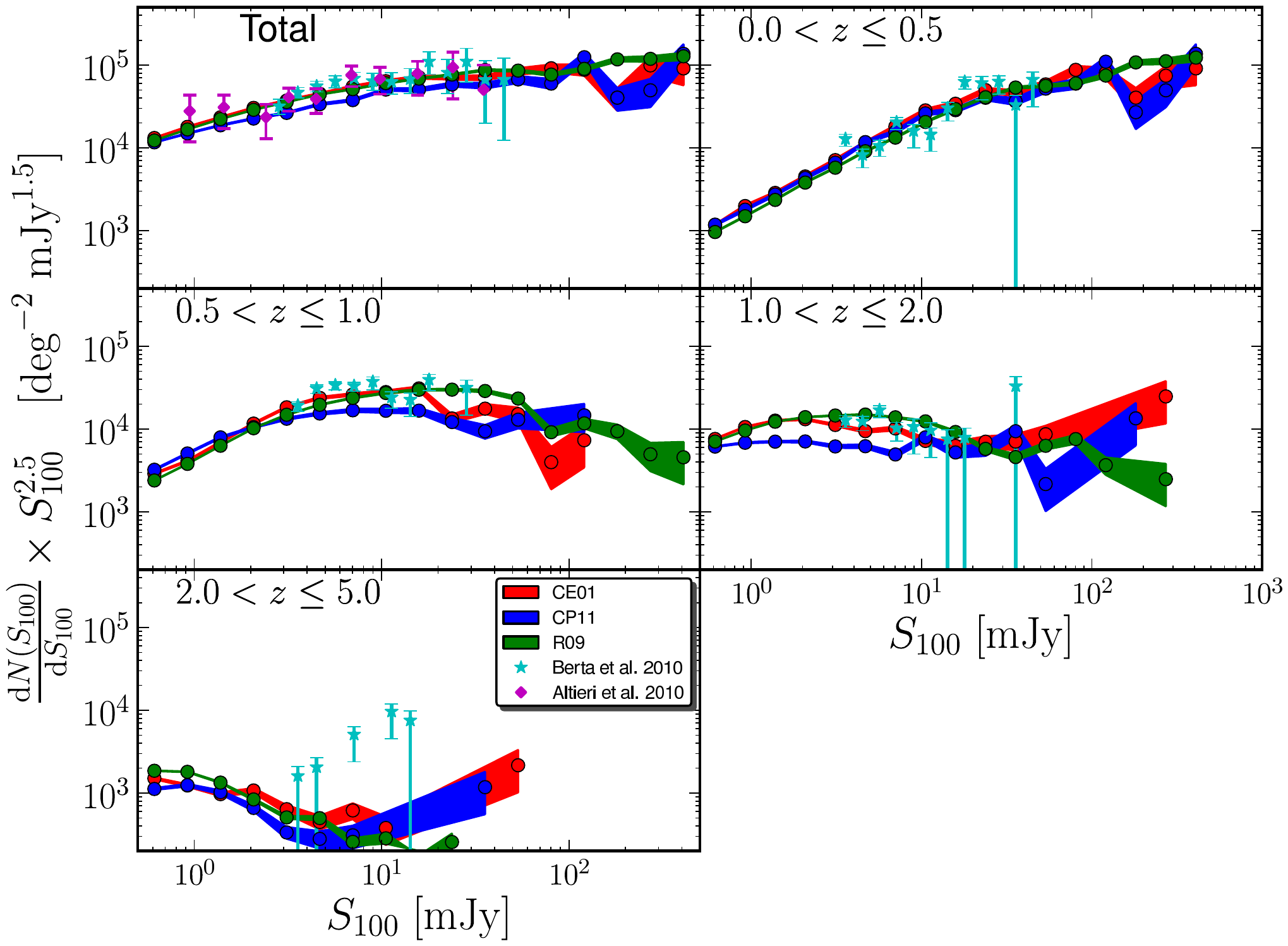}
\caption{Predicted galaxy differential number counts (circles) compared to
observational estimates of \protect\cite{Berta:2010ig} (shown in cyan) and
\protect\cite{Altieri:2010p1044} (shown in magenta) in the PACS 100 \mum band at
different redshift bins. The \protect\cite{Berta:2010ig} observations are based
on direct detections, while the \protect\cite{Altieri:2010p1044} number counts
take advantage of lensing caused by the cluster Abell 2218. The shaded regions
correspond to $3 \sigma$ Poisson errors in the model, while the observational
errors are $1 \sigma$ limits. The different colours refer to different dust
emission template SEDs that have been used in the semi-analytic model (red:
\protect\cite{Chary:2001jf}; blue: \protect\cite{chary-pope:10}; green:
\protect\cite{Rieke:2009p1088} ).}
\label{fig:numbercountsPACS100}
\end{figure}

\begin{figure}
\includegraphics[width=\columnwidth]{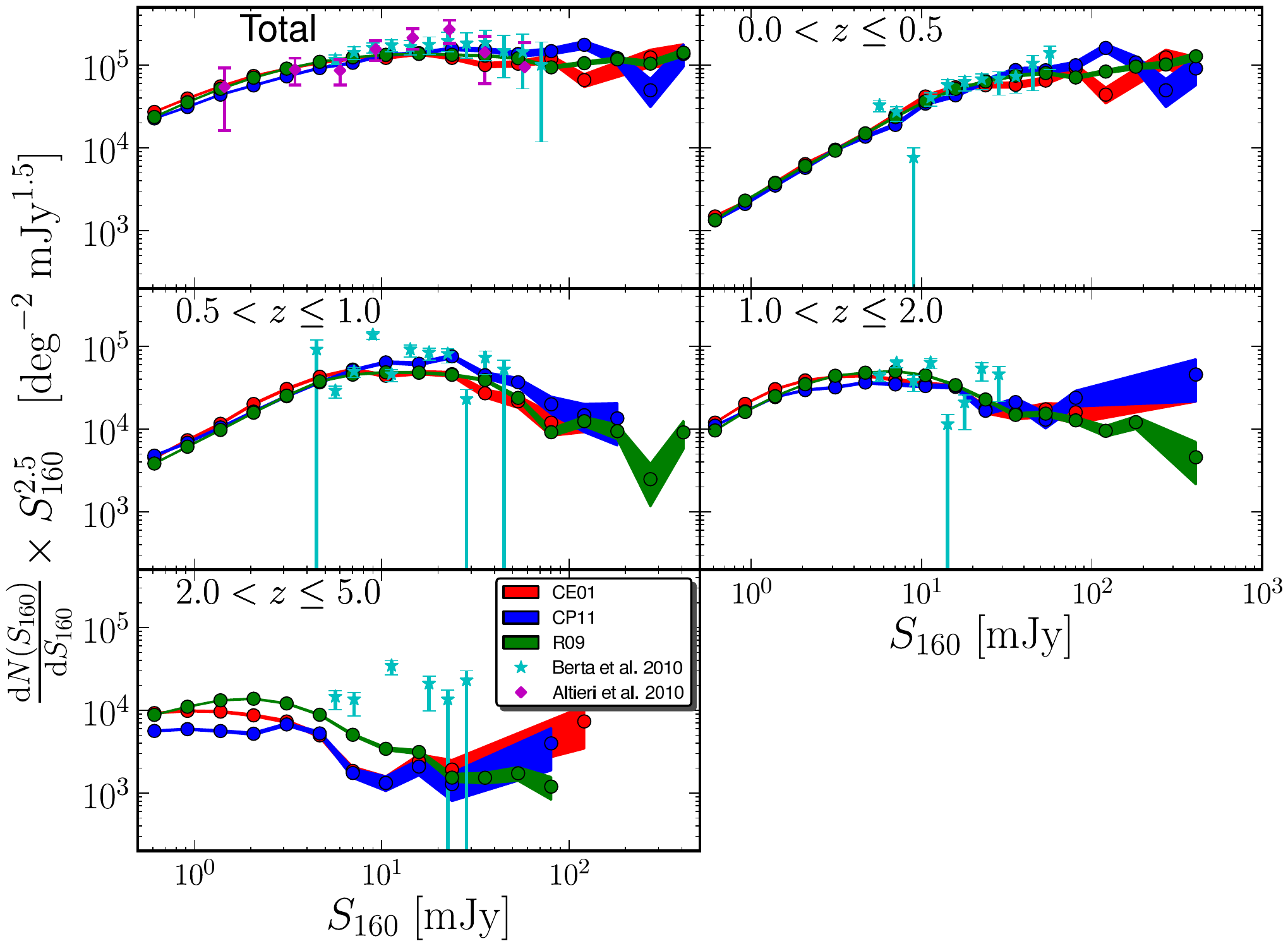}
\caption{Predicted galaxy differential number counts (circles) compared to
observational estimates of \protect\cite{Berta:2010ig} (shown in cyan) and
\protect\cite{Altieri:2010p1044} (shown in magenta) in the PACS 160 \mum band at
different redshift bins. The \protect\cite{Berta:2010ig} observations are based
on direct detections, while \protect\cite{Altieri:2010p1044} number counts take
advantage of lensing caused by Abell 2218. The shaded region corresponds to $3
\sigma$ Poisson errors in the model, while the observational errors are $1
\sigma$ limits. The different colours refer to different dust emission template
SEDs that have been used in the semi-analytic model (red:
\protect\cite{Chary:2001jf}; blue: \protect\cite{chary-pope:10}; green:
\protect\cite{Rieke:2009p1088} ).}
\label{fig:numbercountsPACS160}
\end{figure}

Figures \ref{fig:numbercountsPACS100} and \ref{fig:numbercountsPACS160}
show differential galaxy number counts for the PACS 100 and 160 \mum
bands derived from our simulated light cone after being normalised to a
Euclidean slope $(\mathrm{d}N / \mathrm{d} S \propto S^{-2.5})$. In
order to investigate the sensitivity of our results to the dust emission
templates, we have shown the predictions using three different sets of
templates. The template sets of \citet{Rieke:2009p1088} and
\citet{Chary:2001jf} are based on nearby galaxies. The templates of
\citet{chary-pope:10} incorporate data from distant galaxies, and in
general predict higher fluxes longwards of the thermal dust peak for IR
luminous galaxies, corresponding to colder dust temperatures.
Overplotted are the galaxy differential number counts from the early
\hersch observations of \cite{Berta:2010ig} and
\cite{Altieri:2010p1044}. The number counts of \cite{Berta:2010ig},
which were derived from GOODS-N data, cover the flux range $3-50$ mJy at
$100$ \mum and $5.5-72$ mJy at $160$ \mumn, while the number counts of
\cite{Altieri:2010p1044} cover the flux range $\sim 1 - 35$ and $\sim
1.5 - 58$ mJy at 100 and 160 \mumn, respectively. The general agreement
between the galaxy differential number counts of simulated galaxies and
observations is excellent at low $(z < 2)$ redshift. However, in the
highest redshift bin $(2 < z \leq 5)$ the agreement is worse. Even
though the observational constraints are weaker --- only two (four) of
the observational constraints at the PACS 100 (160) \mum band are not
upper limits --- this disagreement requires further discussion. We
return to this issue later.

A detailed inspection of the differential number counts of the simulated
galaxies shows a monotonic rise in the two lowest redshift bins till $\sim 20$
mJy. This is consistent with a non-evolving population of galaxies. Based on
their observations \cite{Berta:2010ig} argue for a drop at $\sim 20$ mJy, even
though the errors in observations are relatively large. The number counts of the
simulated galaxies provide some support for such a claim at the 100 \mum band,
however, at the 160 \mum band our model does not show any significant drop. For
the PACS 100 \mum band our model predicts a reasonably monotonic rise all the
way to $\sim 500$ mJy. Even so, a small (but statistically significant) drop is
visible at $\sim 60$ mJy. Quantitatively the PACS 160 \mum band's behaviour is
very similar, however, the increase in the Euclidean normalised number counts is
shallower after $\sim 20$ mJy and seems to reach a plateau. The Euclidean
normalised counts seem to rise again at $\sim 200$ mJy, especially at the PACS
100 \mum band.

The redshift evolution of the galaxy differential number counts of the simulated
galaxies in both PACS 100 and 160 \mum bands is rather modest at $z \lesssim 1$.
Even up to redshift two the number of $\sim 10$ mJy IR galaxies seems to evolve
only slowly, in contrast to the number of fainter galaxies, which seems to grow
as a function of redshift, causing the distribution of the number counts to
flatten and eventually to turn over when probing even earlier cosmic times. At
$z \gtrsim 2$ our model predicts that the normalised number counts of the
faintest ($S_{160} \sim 0.1$ mJy) galaxies are at the same level as that of the
brightest ($S_{160} \sim 100$ mJy).

\begin{figure}
\includegraphics[width=\columnwidth]{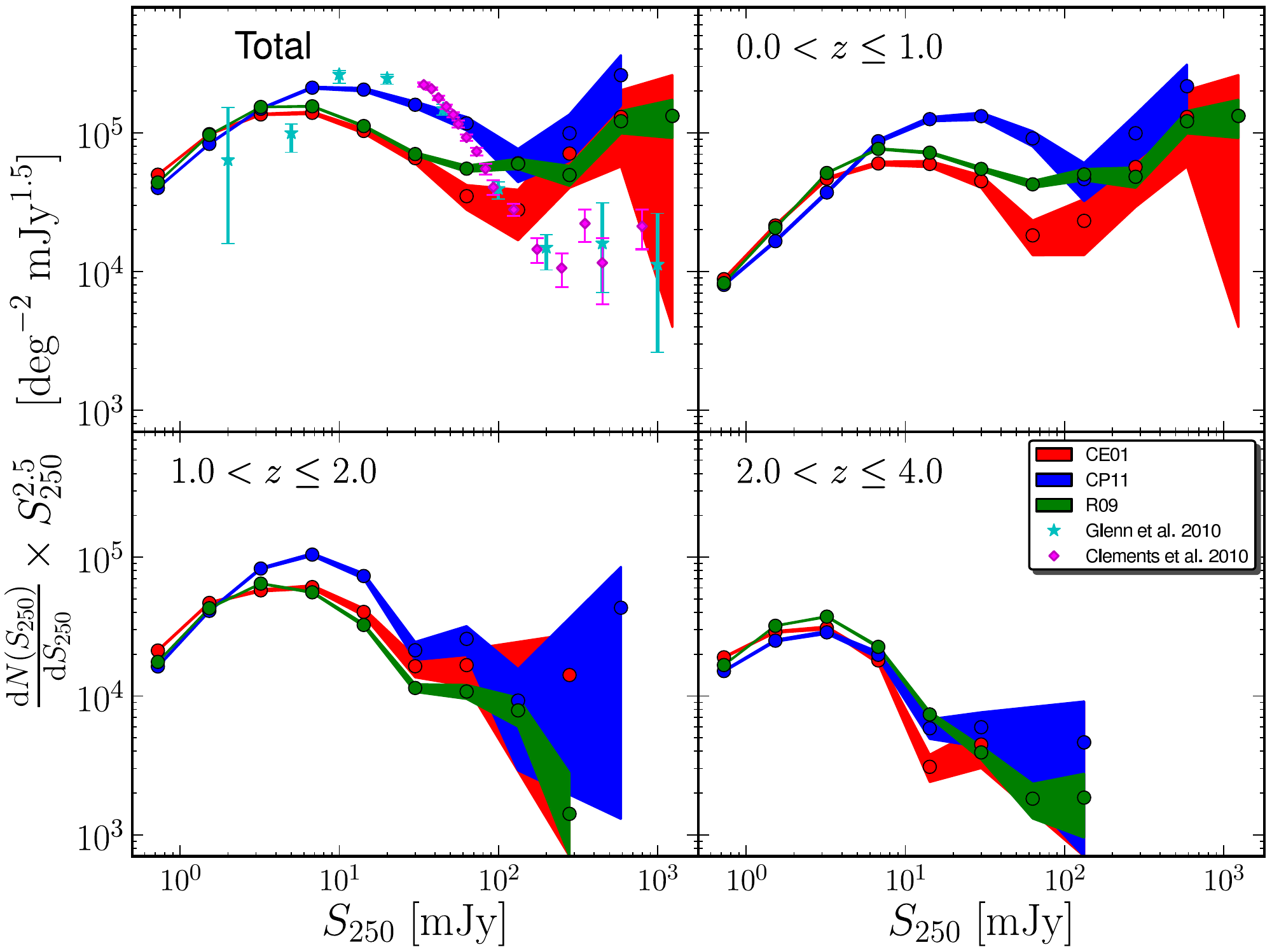}
\caption{Predicted galaxy differential number counts (circles) compared to
observational estimates of \protect\cite{Glenn:2010p1047} (shown in cyan) and
\protect\cite{Clements:2010ce} (shown in magenta) in the SPIRE 250 \mum band in
different redshift bins. The number counts of \protect\cite{Clements:2010ce} are
direct detections, while \protect\cite{Glenn:2010p1047} derived their counts by
modelling the $P(D)$ distribution, rather than identifying individual galaxies.
The shaded regions correspond to $3 \sigma$ Poisson errors in the models, while
the observational errors are $1 \sigma$ limits. Green, red, and blue colours
show the predictions using different dust emission templates, as in the previous
figures.}
	\label{fig:numbercountsSPIRE250}
\end{figure}

\begin{figure}
\includegraphics[width=\columnwidth]{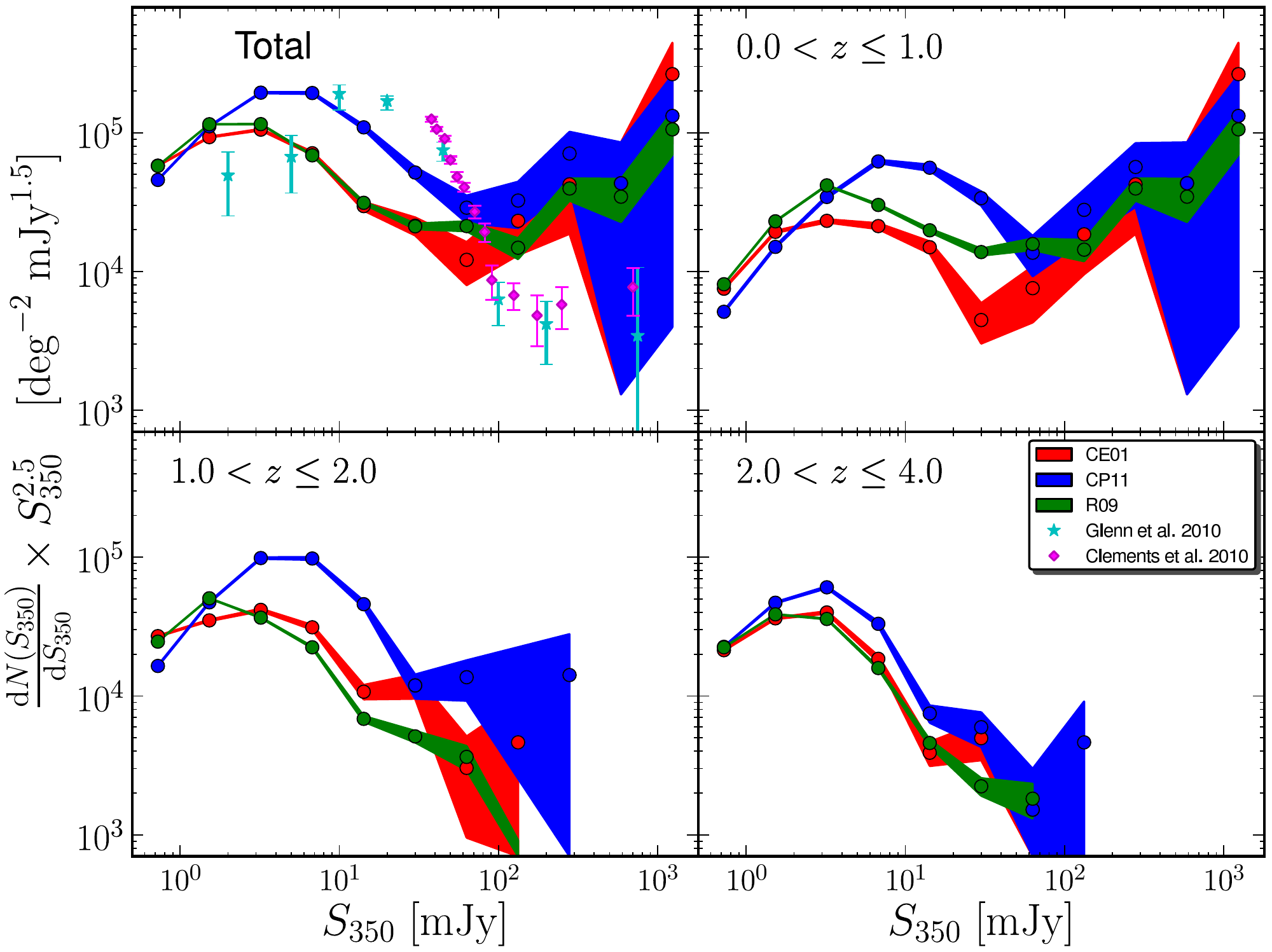}
\caption{Predicted galaxy differential number counts (circles) compared to
observational estimates of \protect\cite{Glenn:2010p1047} (shown in cyan) and
\protect\cite{Clements:2010ce} (shown in magenta) in the SPIRE 350 \mum band at
different redshift bins. The number counts of \protect\cite{Clements:2010ce} are
direct detections, while \protect\cite{Glenn:2010p1047} derived their counts by
modelling the $P(D)$ distribution, rather than identifying individual galaxies.
The shaded regions corresponds to $3 \sigma$ Poisson errors in the models, while
the observational errors are $1 \sigma$ limits. Green, red, and blue colours
show the predictions using different dust emission templates, as in the previous
figures.}
	\label{fig:numbercountsSPIRE350}
\end{figure}

\begin{figure}
\includegraphics[width=\columnwidth]{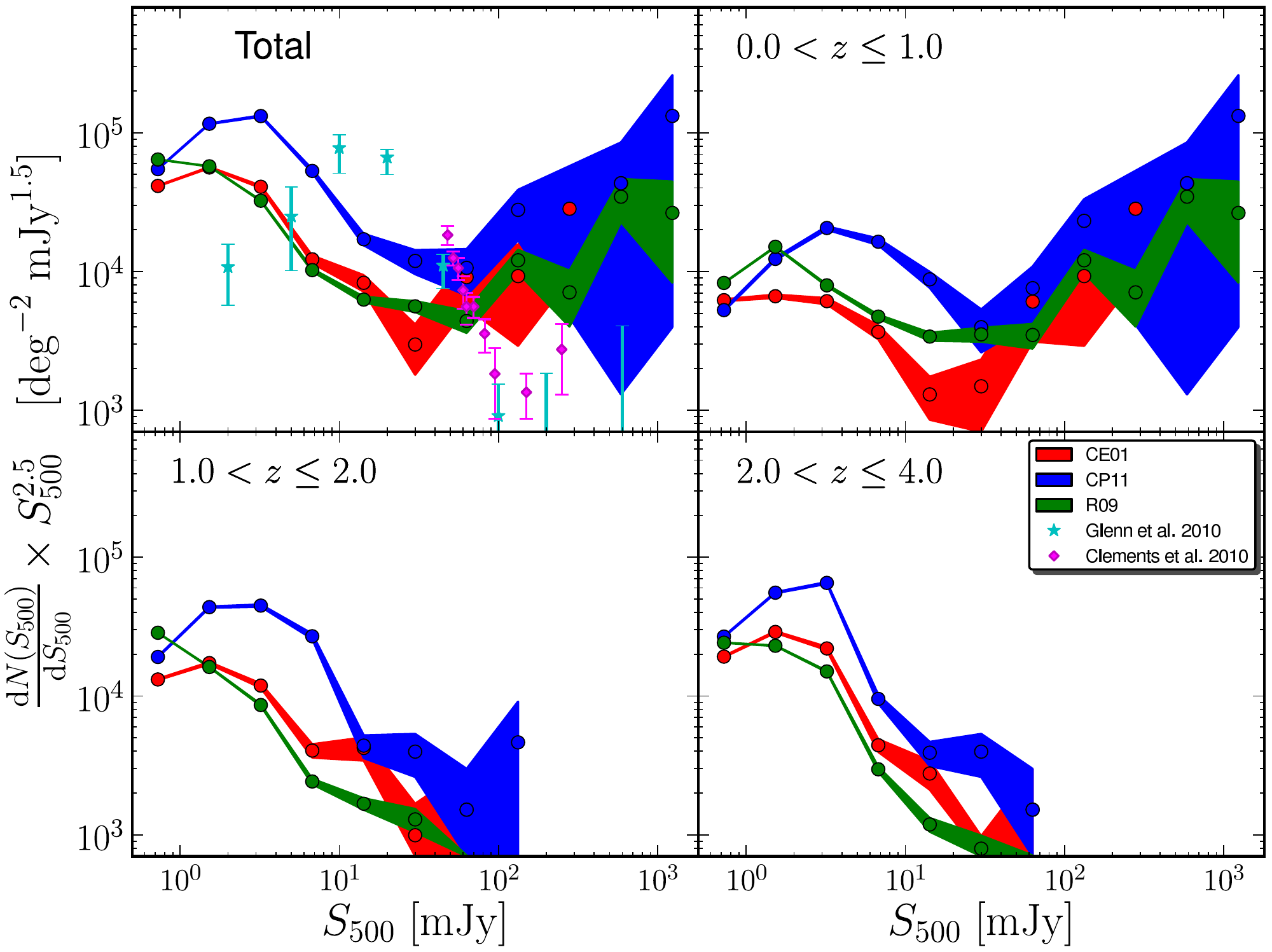}
\caption{Predicted galaxy differential number counts (black circles) compared to
observational estimates of \protect\cite{Glenn:2010p1047} (shown in cyan) and
\protect\cite{Clements:2010ce} (shown in magenta) in the SPIRE 500 \mum band at
different redshift bins. The number counts of \protect\cite{Clements:2010ce} are
direct detections, while \protect\cite{Glenn:2010p1047} derived their counts by
modelling the $P(D)$ distribution, rather than identifying individual galaxies.
The shaded regions correspond to $3 \sigma$ Poisson errors in the models, while
the observational errors are $1 \sigma$ limits. Green, red, and blue colours
show the predictions using different dust emission templates, as in the previous
figures.}
	\label{fig:numbercountsSPIRE500}
\end{figure}

Figures \ref{fig:numbercountsSPIRE250}, \ref{fig:numbercountsSPIRE350}, and
\ref{fig:numbercountsSPIRE500} show Euclidean normalised differential galaxy
number counts derived from our simulated light cone in the 250, 350, and 500
\mum SPIRE bands, respectively. We show the observational results from the early
\hersch observations of \cite{Glenn:2010p1047} and \cite{Clements:2010ce} for
comparison. At 250 \mumn, our model predicts significantly fewer galaxies at the
$5 - 30$ mJy level than are observed. This discrepancy is even larger in the
SPIRE 350 and 500 \mum bands. The agreement seems better at slightly higher
fluxes $(\sim 100$ mJy$)$, but then the models overpredict sources at the
highest flux levels ($\sim 1$ Jy) in all three bands. Note however that the
number of simulated galaxies with $S \gtrsim 400$ mJy is very small, as
indicated by the large errorbars. The actual field-to-field variation will be
even larger, as our model errorbars do not include the effects of large-scale
clustering.

The galaxy differential number count estimates of \cite{Glenn:2010p1047} are
based on the pixel brightness distribution $P(D)$ method, in which no direct
detections are needed. In this method the distribution of pixel brightnesses are
used to make statistical interferences about the slope and amplitude of the
source counts. Therefore, the $P(D)$ method can be used to derive number counts
below the confusion limit (see \cite{Glenn:2010p1047} for details). However, we
note that \cite{Glenn:2010p1047} emphasise that their values are \notice{model\
fits} and thus effectively integral constraints over some region surrounding
each flux density. However, as the differences between their multiply-broken
power-law and spline models are relatively small, we choose to show only their
results based on the multiply-broken power-law model \citep[Table 2
of][]{Glenn:2010p1047}. This provides a reasonable compromise given the fact
that we have added their $1 \sigma$ statistical errors and their estimated
systematic uncertainties in quadrature, after which both of their model fits are
roughly within the errors shown.

\cite{Glenn:2010p1047} note that they clearly detect a break in the number
counts at around $20$ mJy in all the SPIRE bands at high significance. Our model
also shows a break, however, it is at a significantly lower flux ($\sim 2$ mJy),
except for the SPIRE 250 \mum channel when adopting the \cite{chary-pope:10}
templates. \cite{Glenn:2010p1047} also note that there is possible weak $(\sim 1
\sigma)$ evidence for a `bump' in the differential counts around 400 mJy at the
250 \mum band. The location of such a bump in our models is quite dependent on
the adopted dust emission templates, but our model shows an upturn at high
fluxes independent of the templates adopted. However, at the 350 \mum band all
templates show a bump around 300 mJy, although it is unclear whether this is
part of the upturn or a real feature given the relatively large uncertainties at
high fluxes.

We are not aware of any published results for SPIRE counts broken into redshift
bins, however we show our theoretical predictions for comparison with future
observational results. The predicted redshift evolution of the number counts in
all SPIRE bands is qualitatively similar to that of the PACS bands. We note
however that in the SPIRE bands the redshift evolution of LIRGs is more modest.
Even in the highest redshift bin, the Euclidean normalised galaxy differential
number counts of the brightest galaxies are only a factor of a few lower than in
the lower redshift bins.

In summary, our models produce very good agreement for the overall counts in the
PACS 100 and 160 \mum bands, but this agreement rapidly worsens as we move to
longer wavelengths. The overall agreement in the SPIRE bands is quite poor.
Moreover, although our model predictions are also in good agreement with the
galaxy number counts split by redshift for $z < 2$, they do not produce enough
IR-luminous galaxies at $z > 2$. Although the magnitude of the discrepancy
between our theoretical predictions and the observational results is disturbing,
the interpretation is also somewhat complex. In what follows we therefore
discuss potential theoretical and observational complications that could lead to
the noted discrepancies in turn.

We first emphasise that these semi-analytic models have been shown elsewhere
\cite[see S11;][and references therein]{fontanot:09b} to reproduce reasonably
well observed luminosity functions from the rest UV to near-IR, as well as
stellar mass functions and star formation rate functions, from $z \sim 0 - 5$,
although the agreement does get worse towards higher redshifts. Moreover, the
good agreement in the PACS bands up to $z \sim 2$ suggests that many of the
basic ingredients of the model must be reasonable. However, a possible
explanation for some of these discrepancies is that perhaps the adopted dust
emission templates are not valid at earlier cosmic times. Indeed, there is
observational evidence that the correlation between total IR luminosity and dust
temperature might be different in high redshift galaxies \citep{Amblard:2010fn,
Elbaz:2010do, 2010MNRAS.409...75H}. We attempted to investigate this by testing
three different sets of dust emission templates (see also the discussion in
CP11), as described at the beginning of this section. The results of adopting
the different templates are complex, with some templates producing better
results at some wavelengths and redshifts, and others performing better in other
regimes. The predictions with the three templates are similar in the PACS bands
for the overall counts and at low redshifts ($z < 1$). The CP11 templates
predict slightly fewer galaxies in the highest redshift bin ($2 < z < 5$), which
worsens the (already poor) agreement with the observations. However, the CP11
templates produce significantly better agreement for the counts in the SPIRE
bands at intermediate fluxes (between about 3 and 100 mJy).  Thus, we conclude
that the results are sensitive to the choice of dust emission templates, but
none of the template sets that we investigated can reproduce the observations at
all wavelengths. Another potential cause for the noted discrepancy may relate to
our simplified modeling of the dust absorption and emission. We tuned our model
to reproduce the observed ratio of unattenuated to re-radiated light (UV to FIR)
for {\emph average} galaxies at $z=0$. However, galaxies at high redshift may
have more complex geometries, leading to greater scatter or even a breakdown in
this approach. As already noted, scatter or evolution in the dust emission SEDs
could also be important. As noted by CP11, it is not clear whether the
observations they used span the entire range of underlying dust temperatures and
associated far-infrared colours.

On the other hand, one likely culprit in the \hersch observations is the impact
of flux boosting and blending, which affects the counts more severely at longer
wavelengths and may also have a larger effect at high redshifts, where galaxies
are strongly clustered and there is a higher probability that the \hersch beam
intersects lower redshift objects along the line of sight. Indeed, we have
visually checked the Hubble Space Telescope (HST) images at the position of 250
\mum detected \hersch sources in GOODS-N, and it is clear that in many, if not
most cases, there are multiple galaxies within the PACS or SPIRE beam. For
longer wavelengths such as the SPIRE 500 \mum the complications are even worse
because of the full-width-at-half maximum of the beam doubles (from $\sim 18$ at
250 \mum to $\sim 36$ arc seconds). Moreover, according to the simulations
carried out by \cite{2011MNRAS.415.2336R}, more than half ($\sim 57$ per cent)
of sources detected at $\geq 5\sigma$ at $500$ \mum show a flux boosting by a
factor $> 1.5$ and more than every fourth ($\sim 27$ per cent) by a factor $>
2$. This is clearly a very serious issue when comparing theoretical predictions
with these observations.

Other observational complications include cosmic variance. Particularly the
bright counts may be compromised by field-to-field variance, as the areas
covered are still not large. Moreover, at long wavelengths, recent results imply
that many of the bright sources might be lensed
\citep[e.g.][]{2010Sci...330..800N, 2010MNRAS.406.2352L, 2010ApJ...717L..31L}.
Finally, the contribution to the dust heating from obscured active galactic
nuclei (AGN) is uncertain \citep[e.g.][]{Almaini:1999p1158, Granato:1997p1159,
Farrah:2007p1124, Sajina:2008p1126}, although this is unlikely to contribute
significantly particularly at the longer wavelengths. Because we cannot rule out
these effects, for the remainder of this work, we therefore caution that
the predicted IR fluxes that we quote should not be interpreted as corresponding
in a one-to-one fashion with the measured fluxes for individual sources that can
currently be detected in a \hersch image, but rather with the idealised flux
that one would obtain if flux boosting, blending, and cosmic variance
could be overcome. This situation is less than ideal, but we feel that it can
provide a first step towards providing useful predictions. In future work, we
plan to pursue a more rigorous comparison, in which we use the theoretical
simulations to produce mock \hersch images and treat them in the same way as the
real observations. Moreover, the GOODS-\hersch team and others are also working
on a better understanding of the impact of blending on their observations, which
will eventually clarify the interpretation of our predictions. In addition, our
predictions in this form can be compared directly with other theoretical
predictions in the literature, which similarly do not address the effects of
blending.

\subsection{Galaxy Luminosity Functions}\label{ss:luminosity_function}

While simple number counts are the most basic statistic of galaxy populations,
their predictive power is limited. A step further towards the intrinsic
properties of a galaxy population can be taken by studying the galaxy luminosity
functions (LFs). However, even more knowledge of galaxy evolution can be
gathered when the evolution of the rest-frame LFs are studied as a function of
redshift.

\begin{figure*}
\begin{minipage}{18cm}
\begin{center}
\includegraphics[width=17cm]{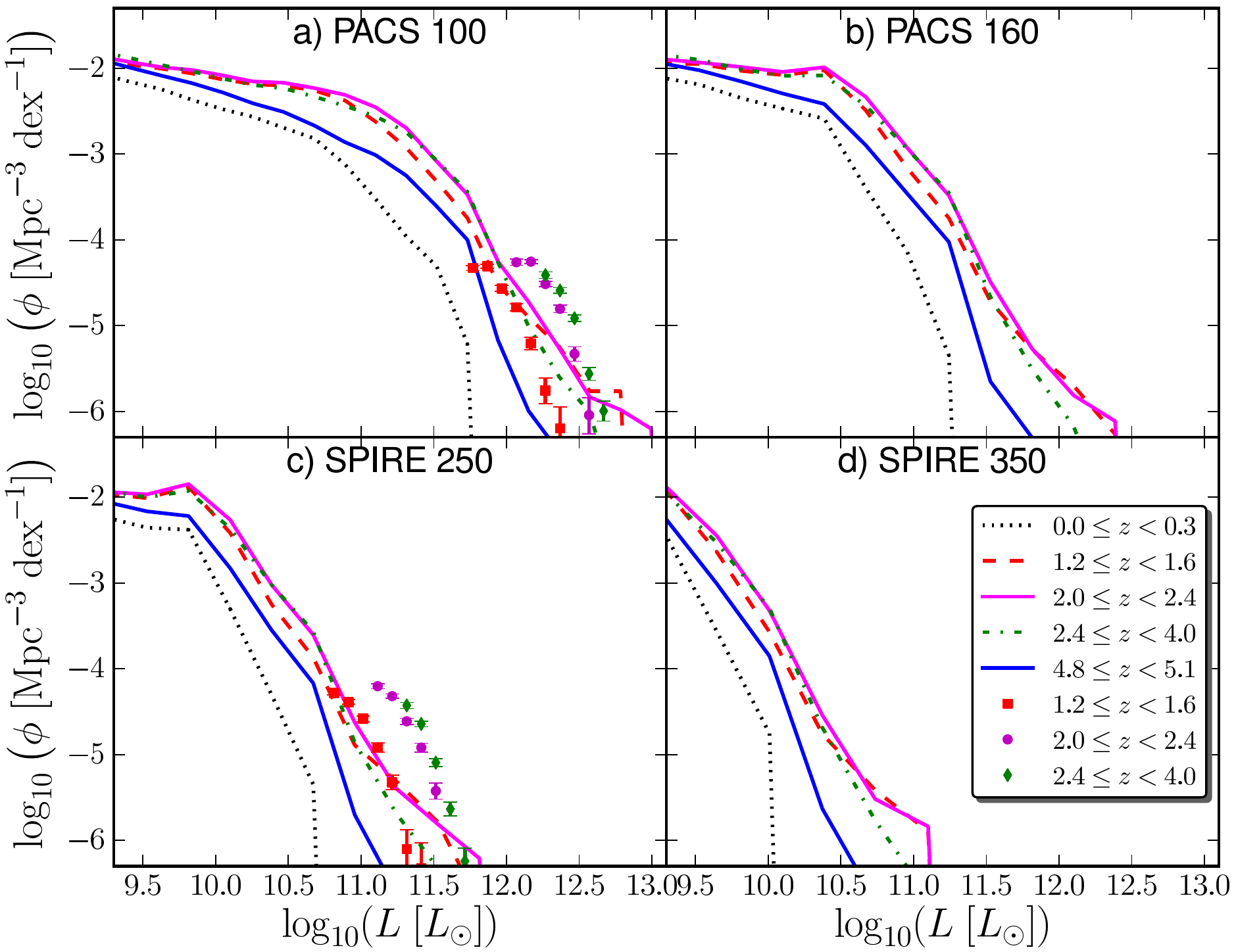}
\caption{Rest-frame galaxy luminosity functions in the PACS 100 and 160 \mum
bands and SPIRE 250 and 350 \mum bands at five different redshifts. Different
redshifts are shown in different colours and are noted in the legend. The
observed luminosity functions (in a and c) marked with symbols are from
\protect\cite{2011arXiv1108.3911L}.}
\label{fig:luminosity_functions}
\end{center}
\end{minipage}
\end{figure*}

Figure \ref{fig:luminosity_functions} shows the rest-frame galaxy LFs in the
PACS 100 and 160 \mum and SPIRE 250 and 350 \mum bands. The rest-frame
galaxy LFs are shown in five different redshift bins: $0 \leq z < 0.3$, $1.2
\leq z < 1.6$, $2.0 \leq z < 2.4$, $2.4 \leq z  < 4.0$, and $4.8 \leq z < 5.1$.
This enables the study of the redshift evolution of the galaxy IR LFs between $z
\sim 5$ and $z \sim 0$. Unfortunately the observational constraints in the
rest-frame LFs of \hersch bands are still limited, especially at higher
redshifts, thus, the comparison between the theoretical predictions and
observations is less than robust. However, to provide an initial comparison we
show the LFs of \cite{2011arXiv1108.3911L}, which have been derived from early
\hersch data. The redshift bins of the theoretical LFs have been matched to the
ones presented in \cite{2011arXiv1108.3911L} to allow easy comparison. However,
the sample of \cite{2011arXiv1108.3911L} is limited to only bright galaxies with
$S_{250} \geq 35$ mJy. Moreover, even though
they estimate that the fraction of false identifications is as low as $\sim 6$
per cent (in areas with good SDSS completeness), they warn that some truly
high-z sources can be missed by their study leading to incompleteness as high as
20 per cent. Such complications will obviously make the comparison less than robust.

The upper panels of Figure \ref{fig:luminosity_functions}, namely a and b,
show the predicted galaxy LFs in the PACS 100 and 160 \mum bands. Qualitatively
these two panels are rather similar, i.e., there are only small differences in
the LFs of the rest-frame 100 and 160 \mum bands. We do note however that the
160 \mum band does not extend to as high luminosities as the 100 \mum band. This
is true at all redshifts. The rest-frame LF of the lowest redshift bin is the
lowest curve, showing that the redshift evolution in the PACS bands is strong.
The two panels show that the number density of the PACS bright galaxies is
highest at redshifts $z \sim 2$, while the highest number density of the faint
galaxies is found at $z \sim 3$. Both PACS bands show that the faint end of the
LFs evolves hardly at all between $z \sim 5$ and $z \sim 2$.  Even though we do
not see much evolution in the faint end of the PACS LFs, a clear difference
exists between the local and higher redshift galaxies. Our model predicts that
the number density of luminous infrared galaxies peaks around $z \sim 2 - 3$, in
agreement with other predictions \citep[e.g.][]{Swinbank:2008eo, Lacey:2010hp}.
Furthermore, the total number of IR bright galaxies is highest at $z \sim 2$.
Such results are easy to understand if the IR rest-frame flux correlates with
dust obscured star formation rate, as the cosmic star formation rate density is
known to peak around the same cosmic time \citep[e.g.][]{Bouwens:2011el}. Our
model also predicts that the number density of the most luminous IR galaxies
(i.e. ULIRGs) increases by a factor of $\sim 5$ between redshifts five and
three, while their number density is found to peak at $z \sim 2$. The evolution
of the PACS LFs also shows that the number density of luminous IR galaxies drops
by a factor of $\sim 10$ between redshifts two and zero. These results emphasise
the strong evolution in the number density of (U)LIRGs as a function of cosmic
time.

In panel a of Figure \ref{fig:luminosity_functions} we also compare our
theoretical LFs to the LFs of \cite{2011arXiv1108.3911L}. Qualitatively similar
redshift evolution is seen in both cases. Our prediction for the lowest redshift
observed LF ($1.2 \leq z < 1.6$) is in reasonable argeement with the
observations, with a modest over-prediction of the highest luminosity sources.
Towards higher redshift, and especially in the highest redshift bin, the
disagreement grows rather large in sense that our model predicts fewer galaxies
with high luminosities than are reported in the observations. Similar effects as
discussed in the case of number counts are likely to contribute. The most
striking difference, however, between our model and the LFs of
\cite{2011arXiv1108.3911L} is the location of the knee of the LFs. Our predicted
LFs turn over at about two orders of magnitude higher number densities. One
potential explanations is provided by \cite{2011arXiv1108.3911L} who argue that
the flattening of their LFs at the lowest luminosities may be, at least in part,
due to the overestimate of the accessible volume yielded by the $1/V_{\rm max}$
estimator for objects near to the flux limit.

The lower panels of Figure \ref{fig:luminosity_functions}, namely c and d,
show the predicted rest-frame galaxy LFs in the SPIRE 250 and 350 \mum bands.
The redshift evolution of the rest-frame SPIRE band LFs is similar to that of
the PACS bands. Again, the number densities of LIRGs and ULIRGs, i.e., the
galaxies at the bright end of the LF, are found to peak at $z \sim 2 - 3$. In
agreement with the PACS bands, the highest number densities are found at
redshifts two and three, while the number densities drop by a factor of $\sim
11$ to redshift zero. Thus, the LFs of the rest-frame SPIRE bands show
qualitatively similar evolution as the LFs of the PACS bands. However, we note
that the bright end of the LFs in the SPIRE bands appears to drop more steeply
than in the PACS bands.

In panel c of Figure \ref{fig:luminosity_functions} we also compare our
theoretical LFs to LFs of \cite{2011arXiv1108.3911L}. As in case of the PACS 100
\mum the faint end of the lowest redshift LFs ($1.2 \leq z < 1.6$) is in good
argeement with our prediction, while in the brightest end our model predicts
more galaxies than are observed. Towards higher redshift, and especially in case
of the highest redshift bin, the disagreement grows rather large.  Again,
similar effects as discussed earlier are likely to play a role.

Infrared LFs have also been derived from other observations
\citep[e.g.][]{Kim:1998p1132, LeFloch:2005p1121, Goto:2010p1062} and
early \hersch data. \cite{Dye:2010p1043} derived SPIRE 250 \mum band
LFs out to $z = 0.5$. They find that the LF exhibits significant
evolution out to $z = 0.5$ and that at a given luminosity, the
comoving space density increases steadily with redshift. The evolution
noted in \cite{Dye:2010p1043} is qualitatively in agreement with our
LFs (see S11), although, their evolution is rather strong given the
rather narrow redshift range. However, strong evolution, particularly
at the bright end, is in agreement with our LFs and also noted by
others \citep[e.g.][]{2009ApJ...707.1779E}. Also
\cite{Vaccari:2010p1049} derived LFs for local galaxies in all three
SPIRE channels. Their LFs are also in good qualitative agreement with
our findings for local galaxies. However, our lightcones have very
small volumes at low redshift, as they were intended for high redshift
comparisons, so a more quantitative comparison is beyond the scope of
this study. A comparison with the $z=0.5$ SPIRE 250 \mum LF is shown in S11.

\section{Physical Properties of IR-Luminous Galaxies over Cosmic Time}\label{s:physical_properties}

Having explored some of the statistical properties of \hersch detected galaxies,
in this Section we present predictions for some of the physical properties of
these objects at different redshifts. We explore the correlations between 160
and 250 \mum flux and many different physical properties of our model galaxies.
Here we show some of the strongest and most interesting correlations that we
identified, which include stellar mass, dark matter halo mass, cold gas mass,
star formation rate, and total IR or bolometric luminosity. The correlations
with the other \hersch bands are similar and all qualitative conclusions would
remain the same, so we do not show them.

\begin{figure*}
\begin{minipage}{18cm}
\begin{center}
\begin{tabular}{cc}
\includegraphics[width=8.8cm]{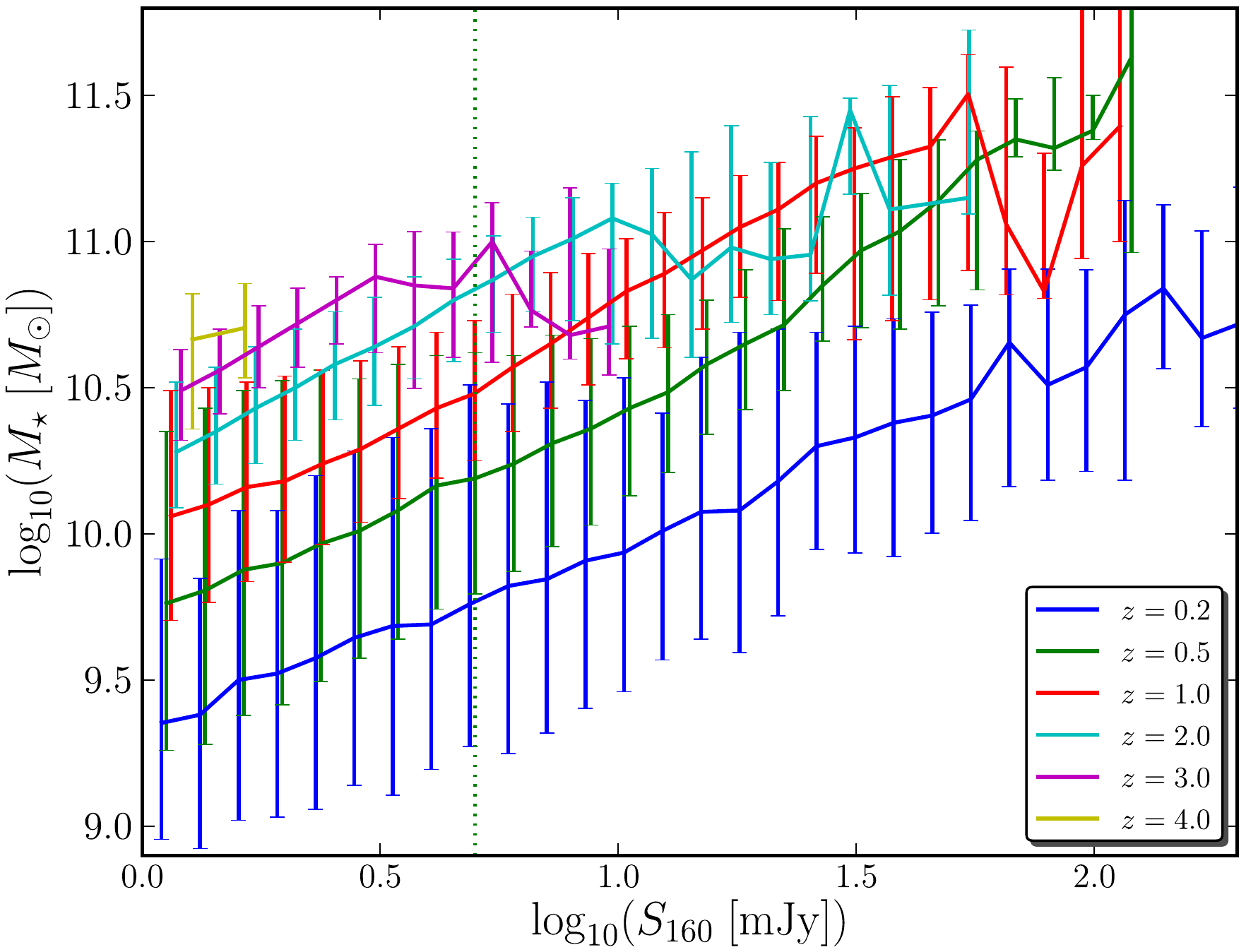} &
\includegraphics[width=8.8cm]{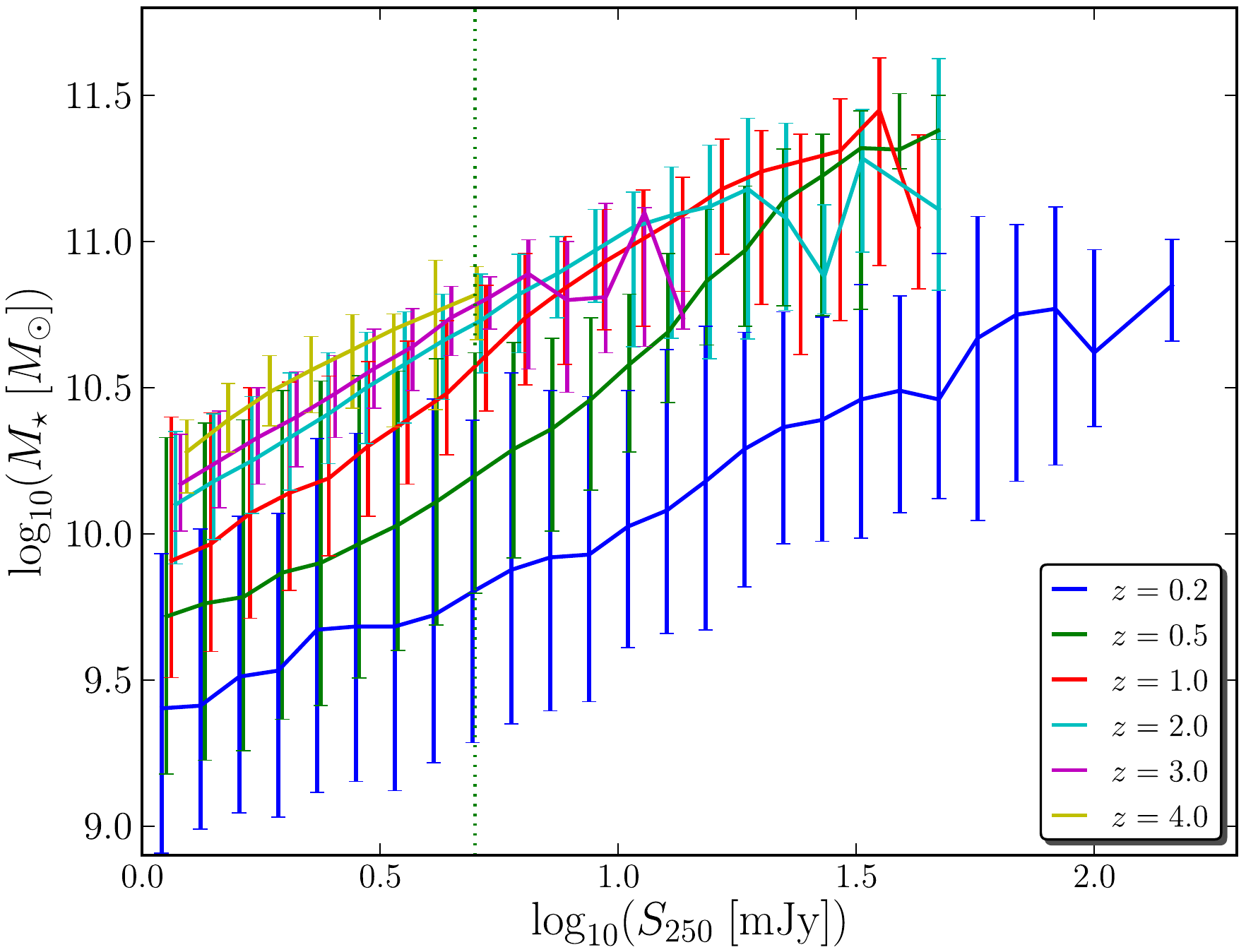}
\end{tabular}
\caption{Predicted median stellar mass vs. flux in the PACS 160 and SPIRE 250
\mum bands for galaxies selected at different redshifts. The green dotted
vertical line shows a representative flux limit of the GOODS-N observations. The
error bars show the 16th and 84th percentiles.}
\label{fig:stellarmassredshift}
\end{center}
\end{minipage}
\end{figure*}

Figure~\ref{fig:stellarmassredshift} shows the predicted median stellar mass as
a function of 160 and 250 \mum flux, along with the 16th and 84th percentiles,
for different redshifts. Our model predicts that all galaxies with $S_{160} > 5$
or $S_{250} >5$ mJy have stellar masses in the range $M_{\star} \sim 10^{9} -
10^{11}$ \msun. Furthermore, we predict that high redshift ($z>2$) galaxies that
can be detected individually in deep \hersch surveys such as the GOODS are quite
massive, $M_{\star} \gtrsim 10^{10.5}$\msun. This is quite different from the
results of \cite{Lacey:2010hp}, who predict much smaller stellar masses for
galaxies at a given FIR flux --- at $z \gtrsim 2$, even their most luminous
galaxies are less massive than $M_{\star} \sim 10^{10}$\msun. This difference
probably arises mainly from the fact that \cite{Lacey:2010hp} adopted a
top-heavy IMF during starbursts. Therefore there are fewer long-lived low mass
stars in galaxies at high redshift. 

\begin{figure*}
\begin{minipage}{18cm}
\begin{center}
\begin{tabular}{cc}
\includegraphics[width=8.8cm]{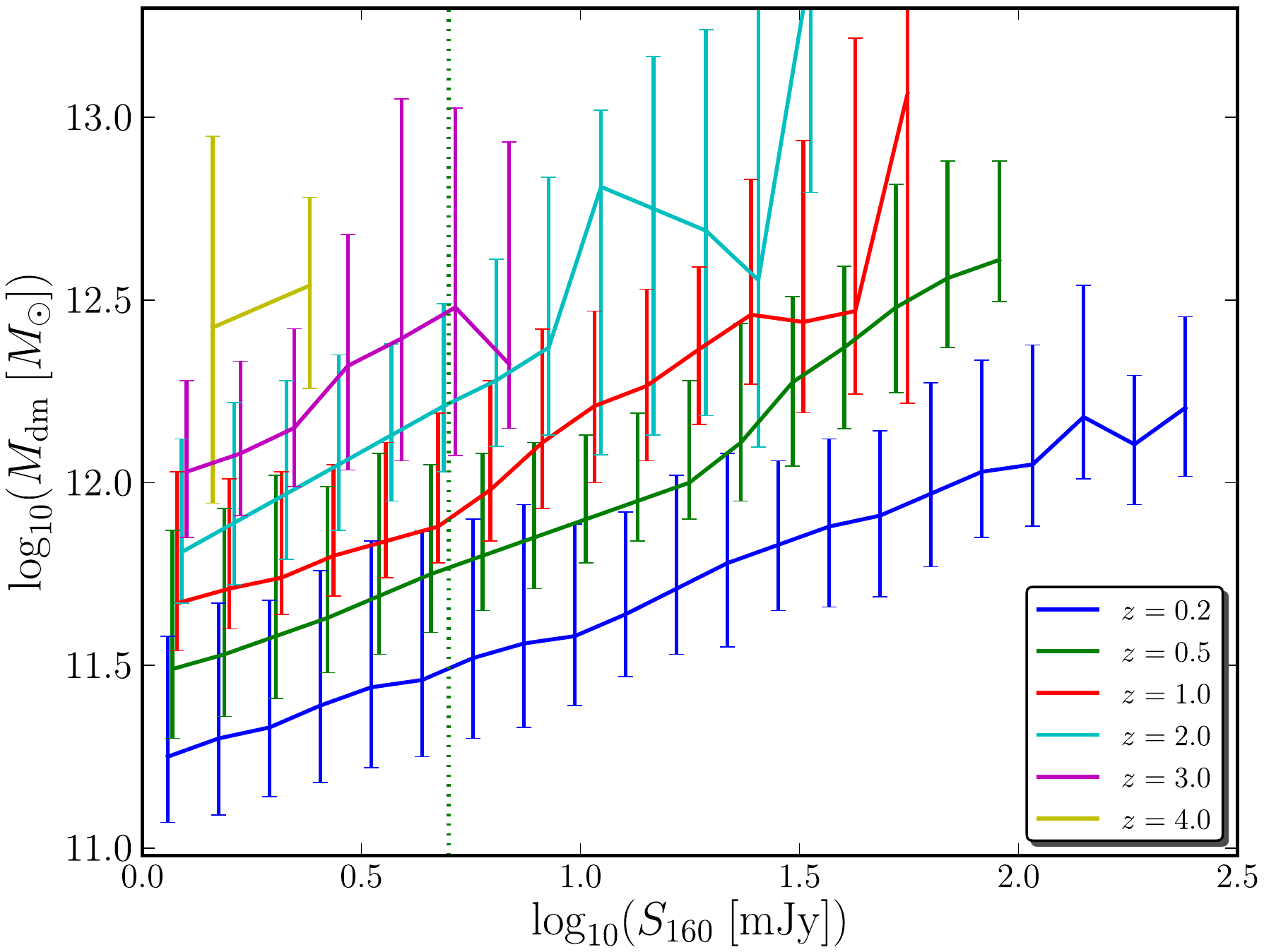} &
\includegraphics[width=8.8cm]{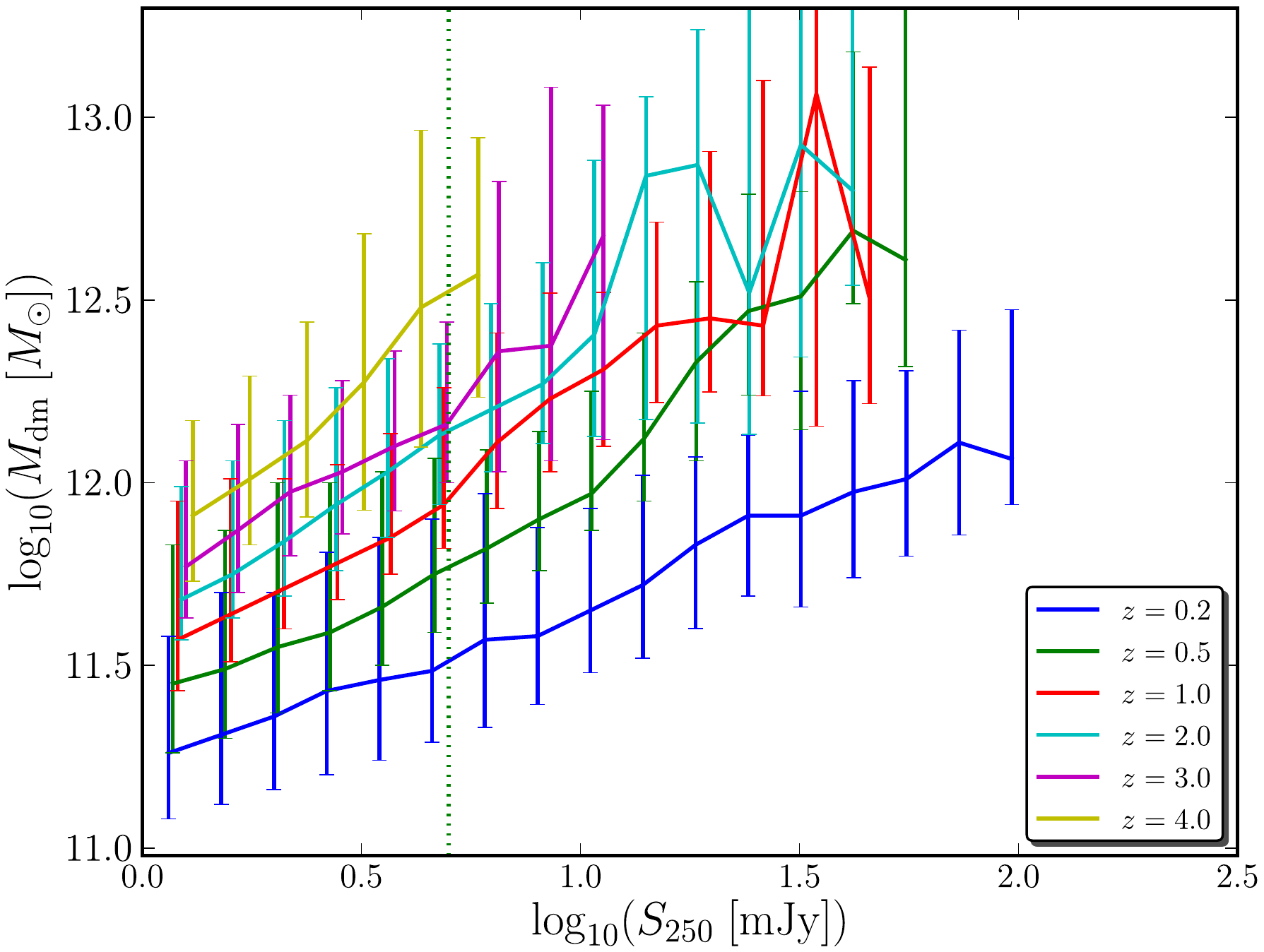}
\end{tabular}
\caption{Predicted median dark matter halo mass $(M_{\mathrm{dm}})$ as a
function of IR flux in the PACS 160 and SPIRE 250 \mum bands at different
redshifts. The green dotted vertical line shows a representative flux limit for
the GOODS-N observations. The error bars show the 16th and 84th percentiles.}
\label{fig:DMmass}
\end{center}
\end{minipage}
\end{figure*}

Not surprisingly, as stellar mass is know to be fairly well correlated with dark
matter halo mass, we find similar results for the dark matter halo masses. At
the lowest redshifts $(z < 0.5)$, luminous IR galaxies (with $S_{160}$ or
$S_{250} > 5$ mJy) can be found in dark matter haloes as light as
$\log_{10}(M_{\mathrm{dm}}/$\msun$) \sim 11.1$ (see Fig. \ref{fig:DMmass}).
However, the bulk of simulated IR bright galaxies with $S_{160}$ or $S_{250} >
5$ mJy, especially at higher redshifts, were found to reside in relatively
massive dark matter haloes. To be more precise, our model predicts that the
masses of the dark matter haloes of IR bright galaxies can cover a broad range
from $\log_{10}(M_{\mathrm{dm}}/$\msun$) \sim 11.5$ to $13.5$, while the bulk of
luminous IR galaxies can be found to reside in rather typical dark matter haloes
with masses $\log_{10}(M_{\mathrm{dm}}/$\msun$) \sim 12.5$.

These predictions are in good agreement with the recent results of
\cite{Amblard:2011p1070}, which were derived from \hersch data. In this paper
the authors find an excess clustering over the linear prediction at arcminute
angular scales in the power spectrum of brightness fluctuations at 250, 350, and
500 \mumn. From this excess \cite{Amblard:2011p1070} infer that IR bright
galaxies are located in dark matter halos with a minimum mass of
$\log_{10}(M_{\mathrm{min}}/$\msun$) = 11.5^{+0.7}_{-0.2}$ at 350 \mumn. When
the authors average over the three wavelengths the minimum halo mass for their
galaxies is at the level of $3 \times 10^{11}$\msun. All our IR bright galaxies
are found to reside in dark matter haloes more massive than this lower limit,
except a handful of local galaxies, which reside in haloes only slightly less
massive ($1 \times 10^{11}$\msun) than the minimum mass of
\cite{Amblard:2011p1070}.

Our simulated galaxies show evidence for a weak correlation between the dark
matter halo mass and the PACS 160 and SPIRE 250 \mum fluxes (Fig.
\ref{fig:DMmass}). Statistically, galaxies residing in more massive dark matter
haloes have higher median IR flux as was also the case with stellar mass (Fig.
\ref{fig:stellarmassredshift}). However, the trend between the dark matter halo
mass and 160 or 250 \mum flux is weaker than in case of stellar mass as
indicated by the larger range between the 16th and 84th percentiles in Fig
\ref{fig:DMmass}. Moreover, for the high redshift galaxies $(2 \leq z < 4)$ with
either $S_{160}$ or $S_{250} > 5$ mJy no statistically significant correlation
can be inferred. Instead, galaxies with the highest fluxes are again found to
reside in dark matter haloes with a broad range of masses from
$\log_{10}(M_{\mathrm{dm}}/$\msun$) \sim 12$ to $13.5$. It is however
interesting to note that luminous IR galaxies can also be found in cluster-mass
dark matter haloes. This implies that some luminous IR galaxies at high redshift
may be the progenitors of present-day brightest cluster galaxies (BCGs).
Moreover, this also implies that the IR bright galaxies at high redshifts may
provide a signpost to identify proto-clusters during the epoch of their
formation. Note, however, that according to our model, the PACS 160 and SPIRE
250 \mum fluxes are almost independent of the dark matter halo mass for
high-redshift galaxies that would be detected in \hersch observations comparable
to fields such as the GOODS. Consequently, the IR flux at high redshift must be
governed primarily by other quantities than the mass of the dark matter halo.

\begin{figure*}
\begin{minipage}{18cm}
\begin{center}
\begin{tabular}{cc}
\includegraphics[width=8.8cm]{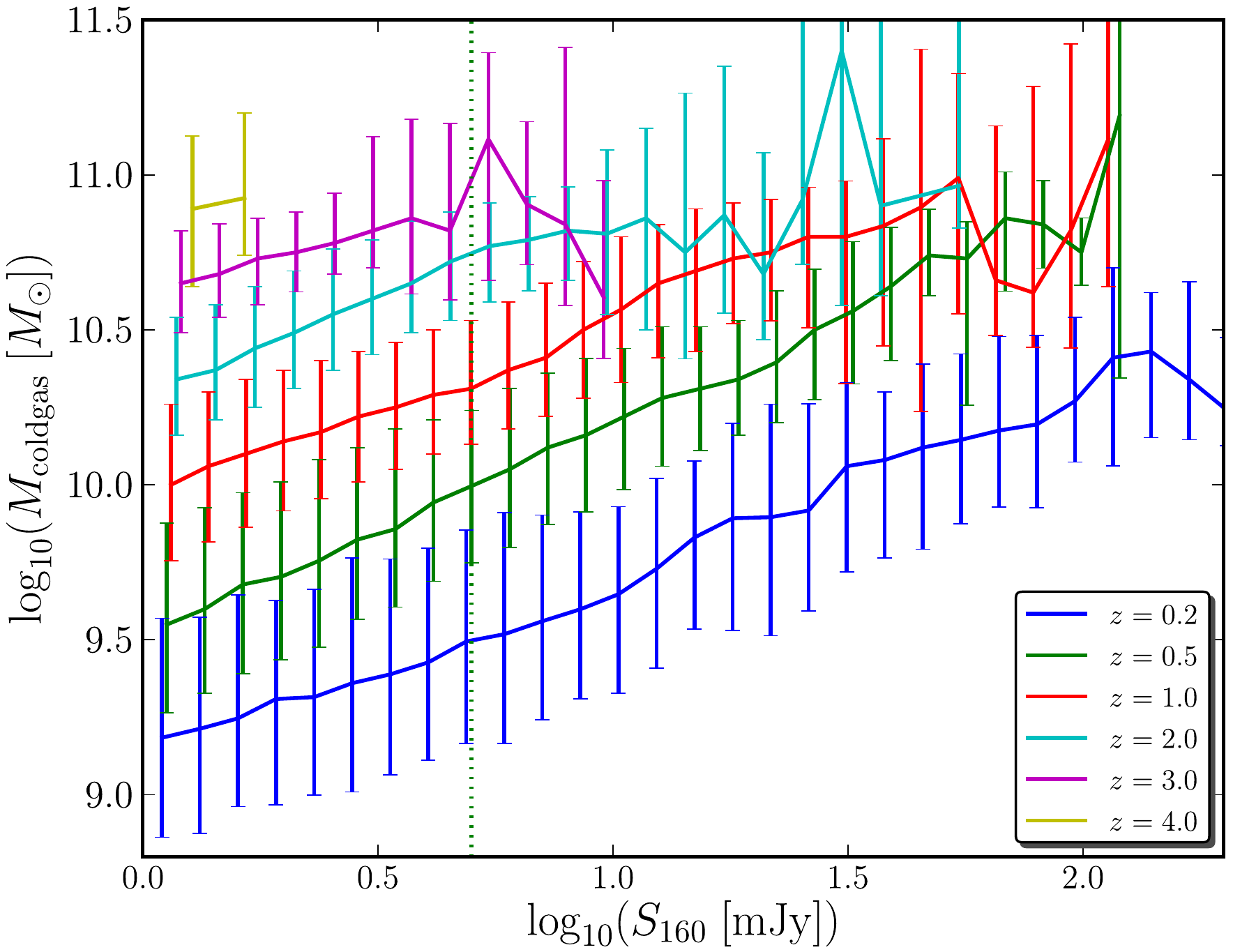} &
\includegraphics[width=8.8cm]{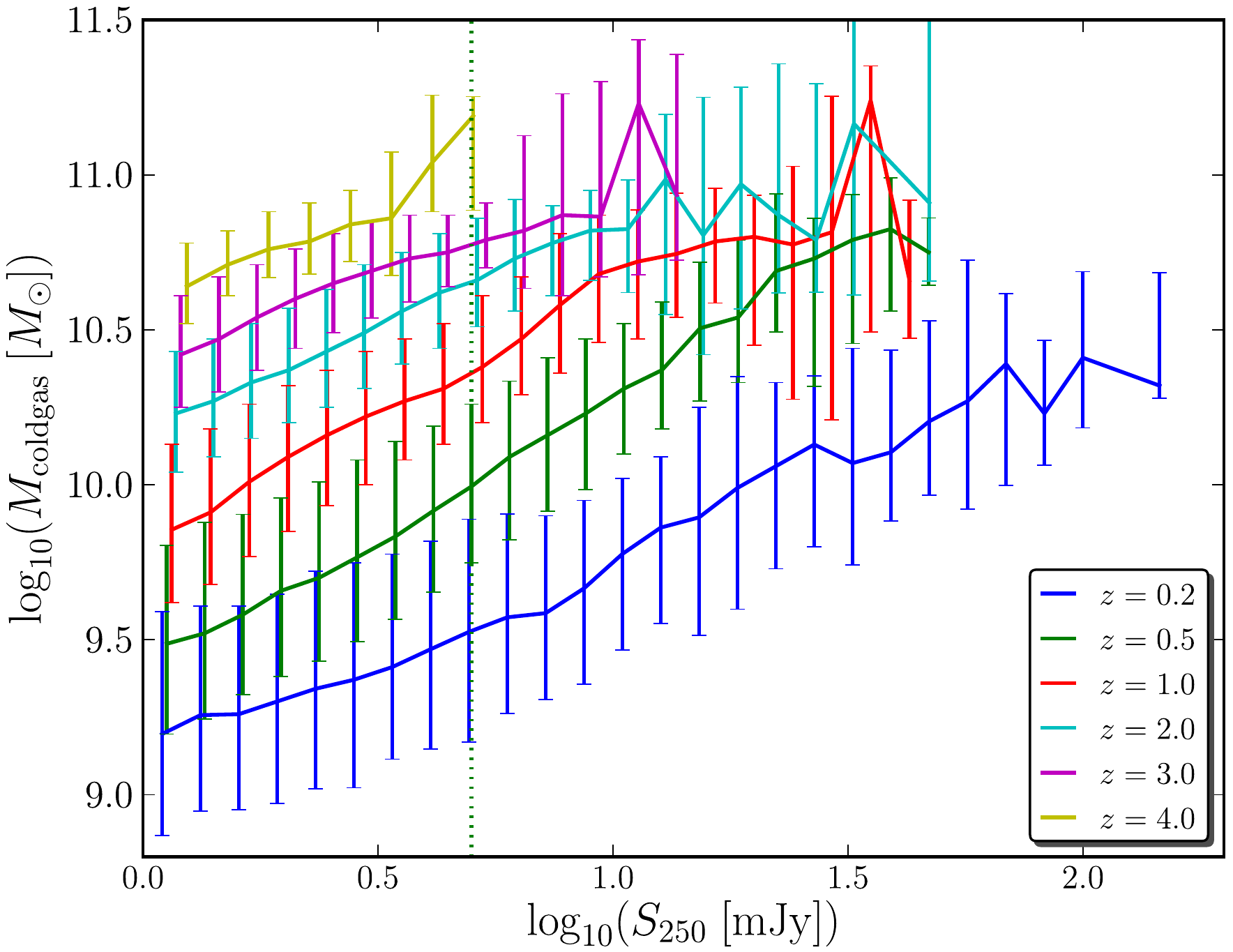}
\end{tabular}
\caption{Predicted median cold gas mass vs flux in the PACS 160 and SPIRE 250
\mum bands for galaxies selected at different redshifts. The green dotted
vertical line shows a representative flux limit of the GOODS-N observations. The
error bars show the 16th and 84th percentiles.}
\label{fig:gasmass}
\end{center}
\end{minipage}
\end{figure*}

Figure \ref{fig:gasmass} shows the predicted median mass in cold gas as a
function of the PACS 160 and SPIRE 250 \mum flux for galaxies selected at
different redshifts. Not surprisingly, our model predicts that at high redshift
the luminous IR galaxies directly detectable in \hersch surveys will be the most
gas rich. Our simulation shows that the galaxies found in these surveys should
typically be gas rich with cold gas masses ranging from $\sim 10^{9}$ up to
$\sim 4 \times 10^{11}$\msun. The bulk of the high redshift IR bright galaxies
were found to contain a cold gas mass of $\sim 6 \times 10^{10}$\msun.
Interestingly, as we noted earlier, the bulk of the high redshift IR bright
galaxies have stellar masses $M_{\star} \sim 6 \times 10^{10}$\msun.
Consequently, the high-redshift luminous IR galaxies should be extremely gas
rich, with the mass in cold gas comparable to or even higher than the mass in
stars for high-redshift luminous IR galaxies. This is a clear prediction for
future observatories such as ALMA, which should be able to detect CO emission
from the molecular gas in these galaxies.

\begin{figure*}
\begin{minipage}{18cm}
\begin{center}
\begin{tabular}{cc}
\includegraphics[width=8.8cm]{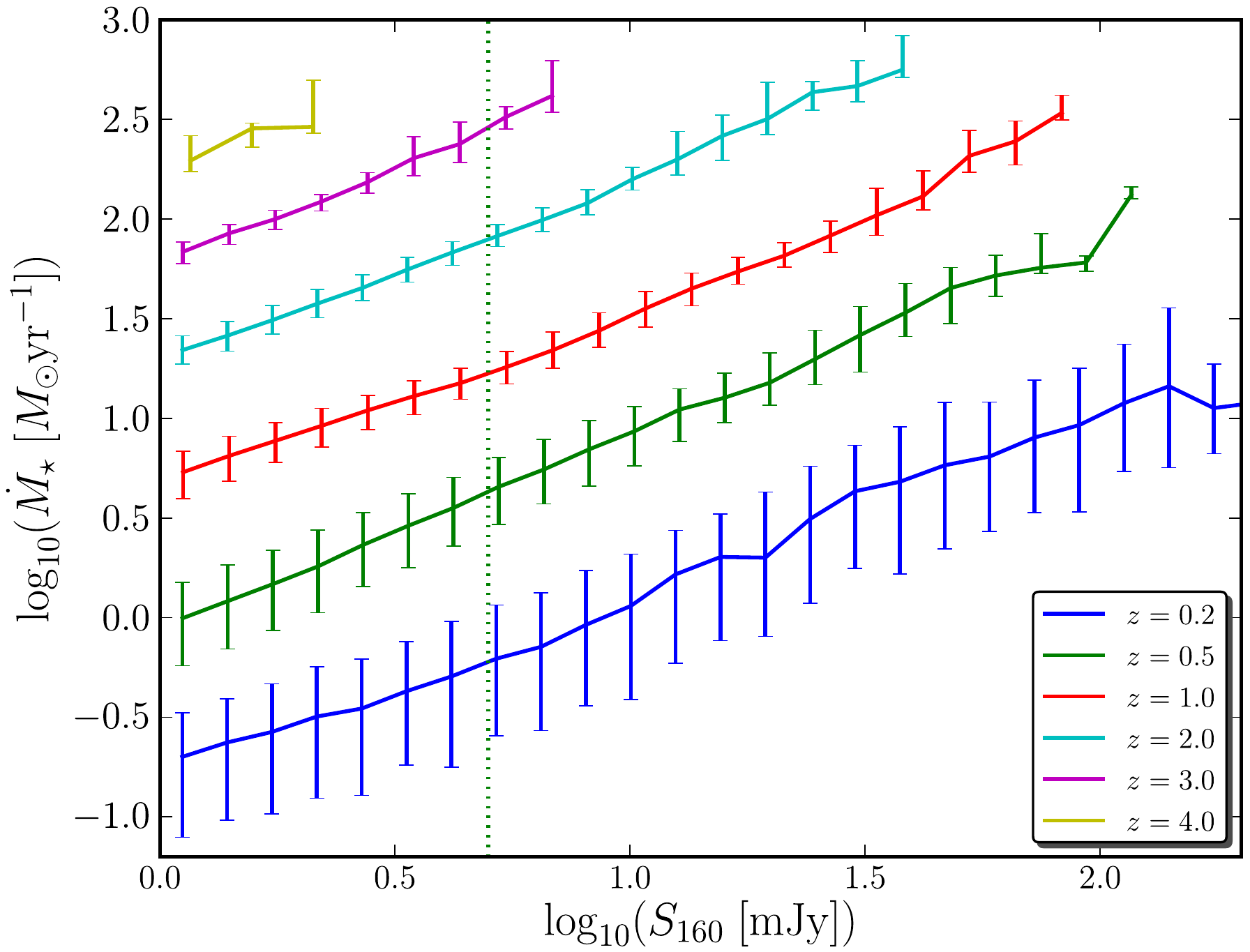} &
\includegraphics[width=8.8cm]{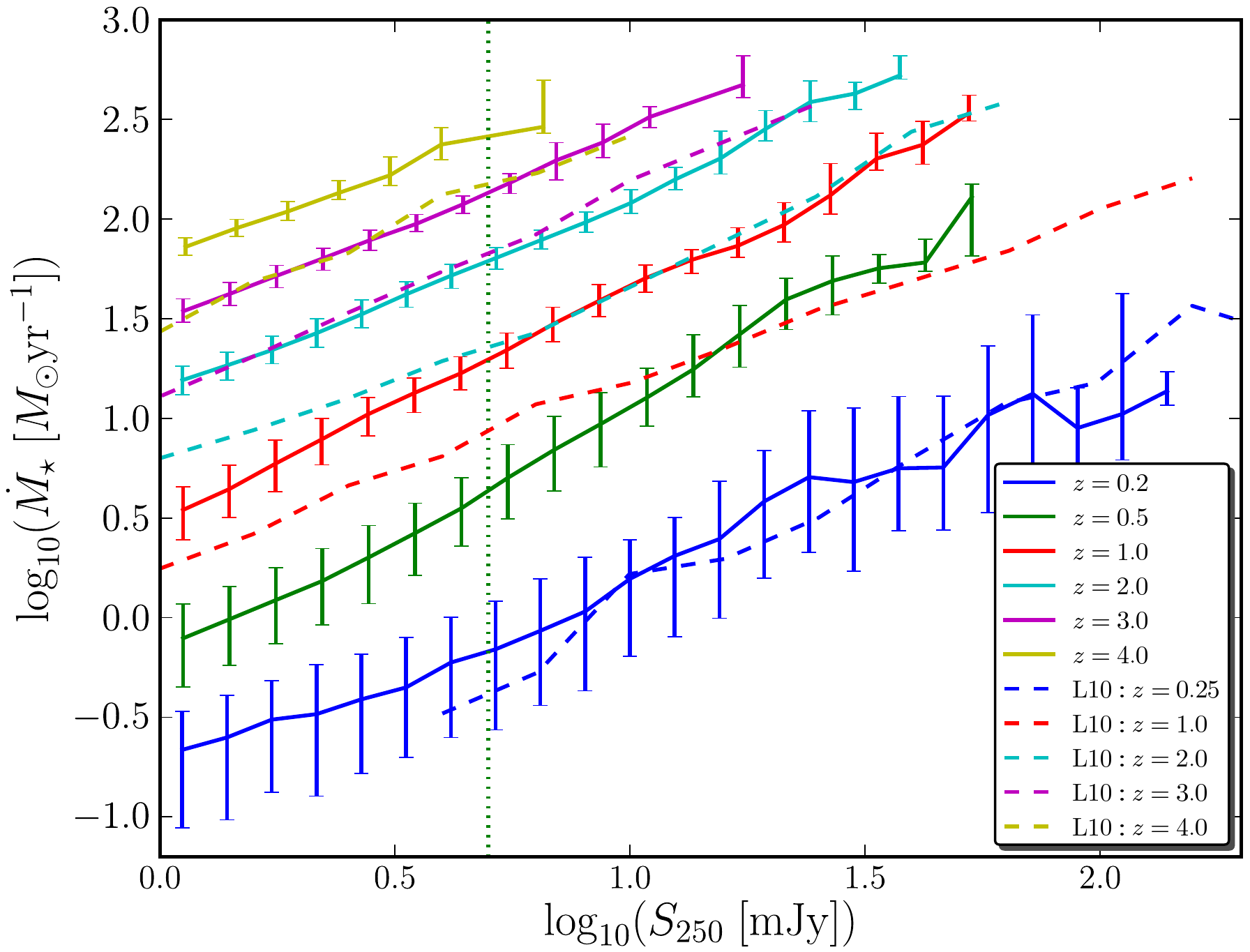}
\end{tabular}
\caption{Predicted median star formation rate as a function of IR flux in the
PACS 160 and SPIRE 250 \mum bands at different redshifts. The green dotted
vertical line shows a representative flux limit for the GOODS-N observations.
The error bars show the 16th and 84th percentiles. The hatched lines on the
right hand side plot (L10) are the predictions from \citet{Lacey:2010hp}.}
\label{fig:sfrIRbright}
\end{center}
\end{minipage}
\end{figure*}

Figure \ref{fig:sfrIRbright} shows the predicted median star formation rate as a
function of the PACS 160 and SPIRE 250 \mum flux at different redshifts. It is
not surprising that this is the strongest correlation seen so far, as the far-IR
flux is generally assumed to be a good proxy for the star formation rate, and at
some level this correlation is artificially tight because of our assumed
one-to-one correspondence between total IR luminosity (light reprocessed by
dust) and dust emission SED. However, this plot provides a convenient guide to
what SFR one expects to be able to power galaxies at a given 160 or 250 \mum
flux at different redshifts, assuming a standard set of templates, as well as an
indication of how much scatter can arise from heating of dust by old stars.

Figure \ref{fig:sfrIRbright} shows that at $z > 2$ the galaxies directly
detectable in surveys such as the GOODS-N should have median star formation
rates $> 30$\msun yr$^{-1}$ according to our model prediction. We find that the
average star formation rate for luminous IR galaxies at a high redshift $(2 \leq
z < 4)$ is $\sim 155$\msun yr$^{-1}$. This is significantly higher than for all
galaxies (with $\log_{10}(M_{\mathrm{dm}}) > 9$) in the same redshift range, for
which our model predicts a mean star formation rate of merely $\sim 2$\msun
yr$^{-1}$. The highest star formation rate predicted by our model for a
high-redshift galaxy is as high as $\sim 7800$\msun yr$^{-1}$. In the next
Section, we explore the physical mechanisms that are driving high SFRs in high
redshift galaxies. Clearly, such high SFR cannot persist for very long periods
of time!

\begin{figure*}
\begin{minipage}{18cm}
\begin{center}
\begin{tabular}{cc}
\includegraphics[width=8.8cm]{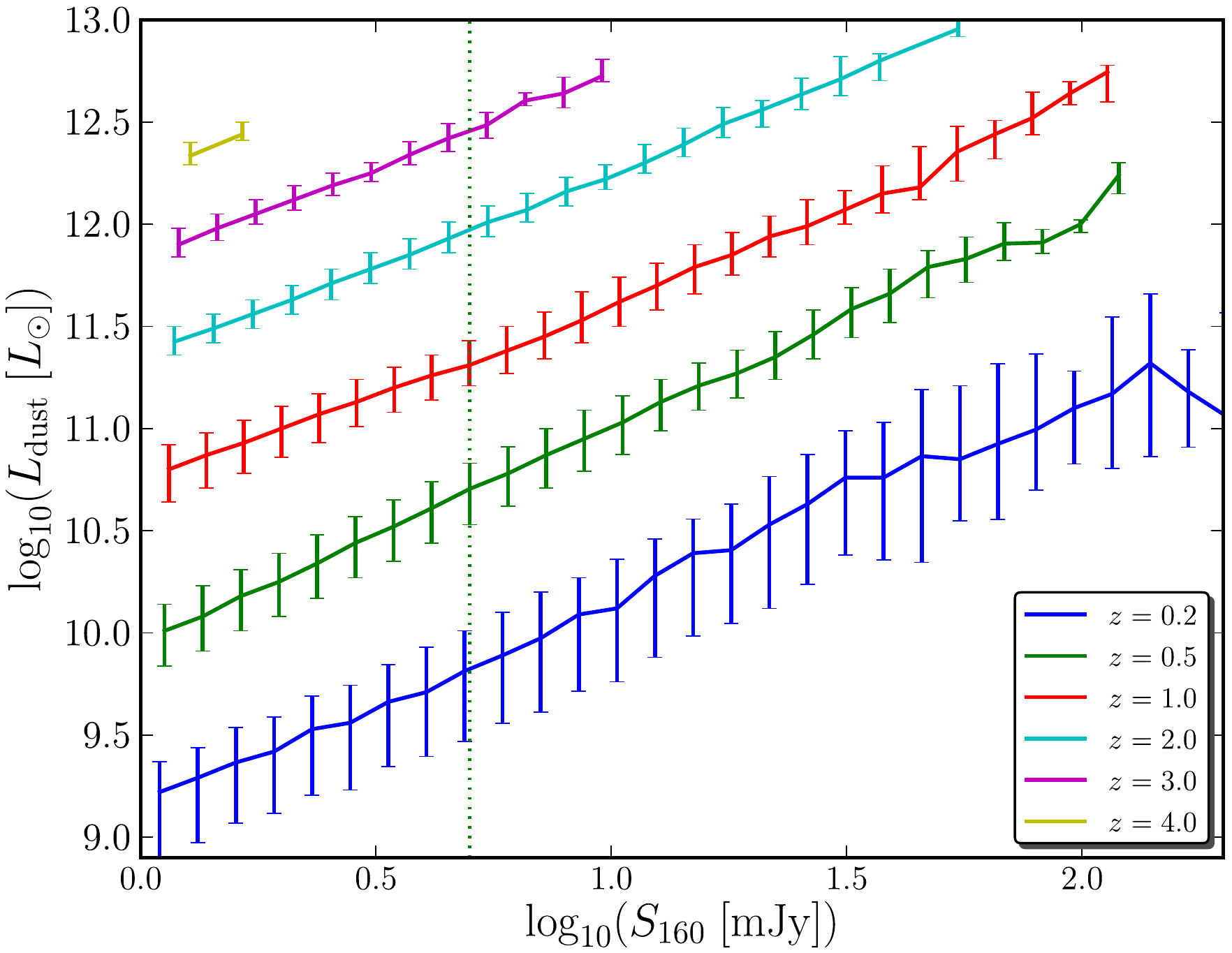} &
\includegraphics[width=8.8cm]{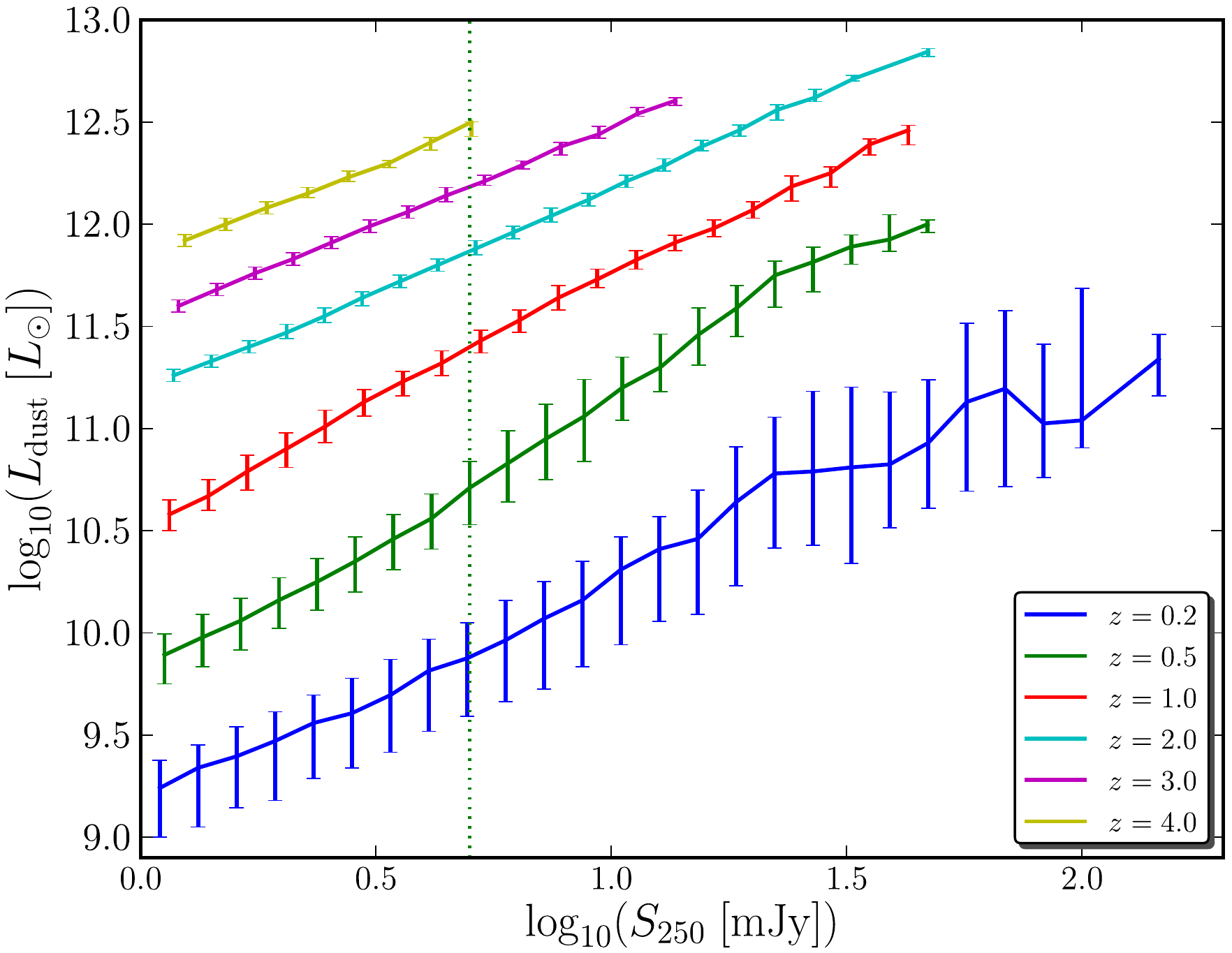}
\end{tabular}
\caption{Predicted median total IR luminosity $(L_{\mathrm{dust}})$ as a
function of IR flux in the PACS 160 and SPIRE 250 \mum bands at different
redshifts. The green dotted vertical line shows a representative flux limit for
the GOODS-N observations. The error bars show the 16th and 84th percentiles.}
\label{fig:totalIR}
\end{center}
\end{minipage}
\end{figure*}

Figure \ref{fig:totalIR} shows the median values for the total IR luminosity
$(L_{\mathrm{dust}})$ as a function of flux in the PACS 160 and SPIRE 250 \mum
bands at different redshifts. This plot illustrates that, at low redshift, only
a very few of the brightest galaxies in our (relatively small volume)
simulations would not even be considered LIRGs ($L_{\mathrm{IR}} >
10^{11}L_{\odot}$). Our simulations do not contain any ULIRGs ($L_{\mathrm{IR}}
> 10^{12}L_{\odot}$) at low redshift. At $z>1$, all galaxies above our nominal
GOODS-\hersch detection limit would be considered LIRGs or ULIRGs, at $z\sim2$
all of these galaxies would qualify as very bright LIRGs or ULIRGs, and at $z >
3$ all of these galaxies are ULIRGs.

At high redshift (\redshift) our model predicts median SFRs that are a factor of
two higher than those reported in \cite{Lacey:2010hp}, as show in Fig.
\ref{fig:sfrIRbright}. As already noted, \citet{Lacey:2010hp} assumed a
top-heavy IMF in starbursts. With a top-heavy IMF, in a given star formation
episode, a larger fraction of the mass goes into massive stars that can
efficiently heat the dust. Thus, for a top-heavy IMF, the relationship between
SFR and IR luminosity is shifted such that a given SFR results in higher IR
luminosities. In the \cite{Lacey:2010hp} model, the fraction of star formation
occurring in the burst mode increases with redshift, so the average IMF with
which stars are being formed shifts from being close to a solar neighbourhood
IMF at the present day to being very top-heavy at high redshift. This helps to
explain why our SFR vs 250 \mum flux relationship agrees with that of
\cite{Lacey:2010hp} in the local Universe $(z < 0.25)$, but differs
significantly at earlier cosmic epochs. In addition, the treatment of dust is
also different in the \cite{Lacey:2010hp} model. They used results from the
RT+dust model calculations of \cite{silva:98}.

\subsection{Sizes of High-redshift Late-type Galaxies}\label{ss:sizes}

\begin{figure}
\includegraphics[width=\columnwidth]{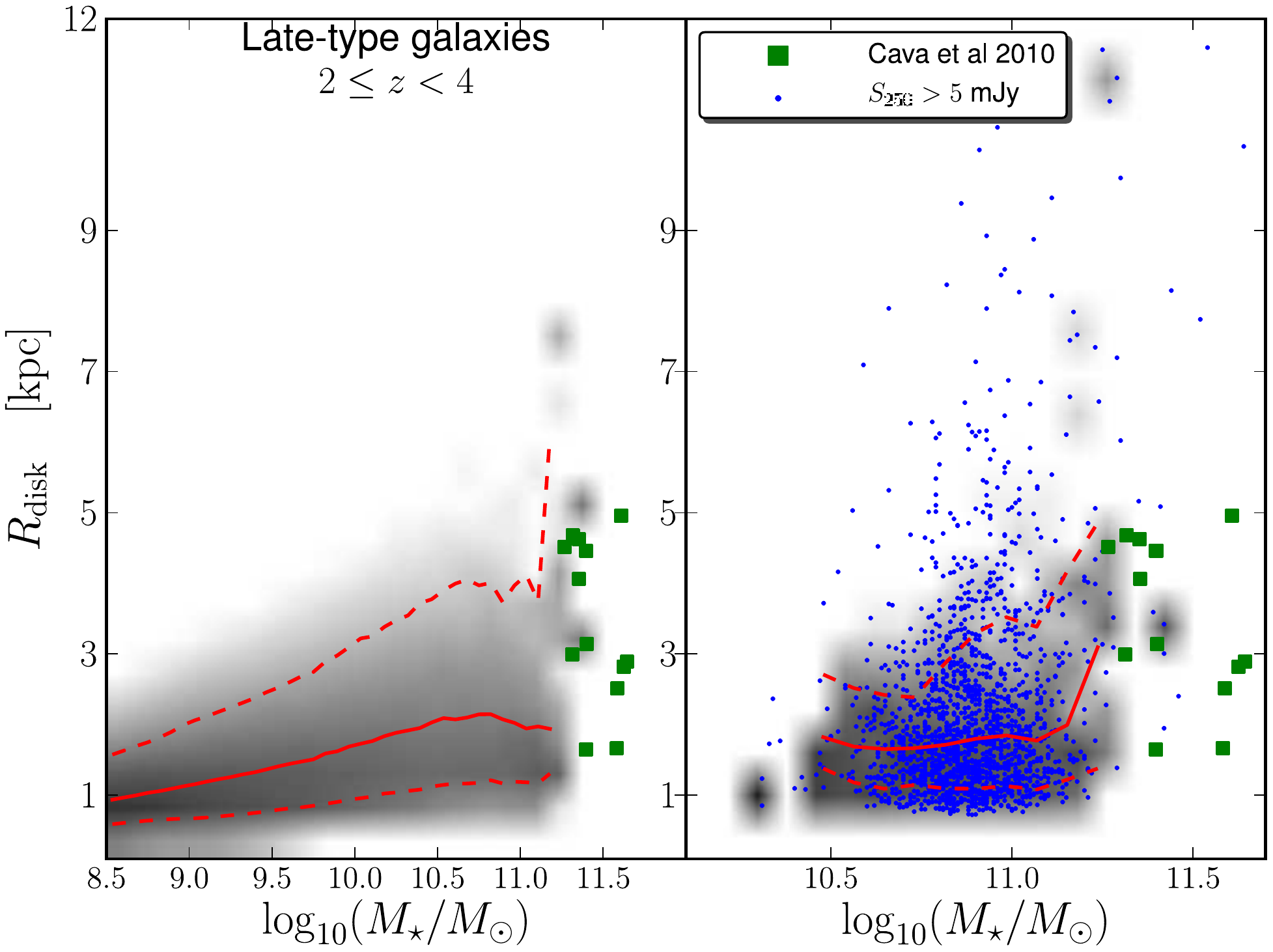}
\caption{Hess plot of disk sizes $(R_{\mathrm{disk}})$ of the late-type galaxies
as a function of stellar masses $M_{\star}$. All galaxies shown are in the
redshift range \redshift\ and have not experienced a merger within the last 500
Myr. The right-hand panel shows only late-type galaxies that are brighter than 5
mJy in the SPIRE 250 \mum band. The red lines show the median and $16$ and $84$
percentiles for simulated galaxies. The observed late-type galaxies, shown with
green squares, are from \protect\cite{Cava:2010p1050}.}
	\label{fig:size}
\end{figure}

For this sub-section we use our ``high-redshift'' SPIRE 250 \mum
selected sample of galaxies selected from our lightcones in the
redshift range \redshift. We identify disk galaxies using the bulge to
stellar mass ratio: galaxies with $M_{\mathrm{bulge}} /
M_{\mathrm{total}} \leq 0.4$ are considered as late-type galaxies. We
also removed all galaxies that have experienced one or more mergers
within the last $500$ Myr, because these galaxies have disturbed disks
(and morphologies) and thus the size of the stellar disk is
ill-defined.  Figure \ref{fig:size} shows a Hess plot of the galaxy's
disk size as a function of its stellar mass $M_{\star}$ in physical
units for this sample of high-redshift galaxies. For a comparison, we
also plot the sizes and stellar masses from \cite{Cava:2010p1050} with
green squares. The gray shading in both panels of
Fig. \ref{fig:size} describes the number of simulated galaxies in a
given two dimensional bin, while the red lines show the median and
$16$ and $84$ percentiles.

As in the real Universe, our model disks show a positive correlation
between stellar mass and radial size. Very few galaxies are predicted
to be as large in radial extent as the galaxies presented in
\cite{Cava:2010p1050}. However, the right panel of
Figure~\ref{fig:size} shows that when we select galaxies in the same
redshift range with $S_{250} > 5$ mJy, the predicted sizes are in good
agreement with the observational results. The late-type galaxies in
our $S_{250} > 5$ mJy high-redshift sample have average disk radii of
$\sim 2.2$ kpc. This is significantly larger than the mean disk size
($\sim 0.9$ kpc) of all late-type galaxies in the same redshift range
(\redshift). However, the small mean disk size in case of all
late-types is driven by the large number of dwarf galaxies. But even
if we use the same stellar mass cut that the SPIRE 250 \mum band
selection induces, we find that the average size of the stellar disks
is almost a factor of two smaller $(\sim 1.3$ kpc$)$ than the mean
disk size of the high-redshift sample. Consequently, in our models,
the IR bright late-type galaxies that have not experienced a merger
within the last $500$ Myr are on average larger than galaxies that
emit lower SPIRE 250 \mum fluxes. We speculate that this is because
our models predict that larger galaxies are more gas rich, and we have
already seen that there is a correlation between gas mass and IR
luminosity. Our high-redshift sample of luminous IR galaxies also
contains a few late-type galaxies whose stellar disks are $> 7$ kpc in
size \citep[for similar findings, see][]{Rujopakarn:2011bh}.

\subsection{Correlation with Merger Activity}\label{ss:mergers}

Ultra-luminous galaxies in the local Universe are widely believed to be caused
by major galaxy mergers \citep[e.g.][]{Sanders:1996p1085, Colina:2001p1123,
Farrah:2003p1149, Dasyra:2006p1119, Vaisanen:2008p1105}. It has also been argued
that ULIRGs will evolve into moderate-mass ellipticals
\citep[e.g.][]{Genzel:2001p1106, Tacconi:2002p1129, Dasyra:2006p1119} and/or AGN
\cite[][and references therein]{Farrah:2003p1149} after the merger or
interaction. Merger driven star formation has also been studied using
hydrodynamic simulations \citep[e.g.][]{Jog:1992p1118, Mihos:1996p1093,
Cox:2008p1141, Teyssier:2010ia}, which support the picture that massive
starbursts could be driven by mergers.  It has been argued
\citep[e.g.][]{Dasyra:2006p1136} that the majority of ULIRGs are triggered by
almost equal-mass major mergers and that less violent mergers of mass ratio
$<1:3$ typically do not force enough gas into the nucleus to generate ULIRG
luminosities. This is in agreement with simulations (e.g. \cite{Cox:2008p1141},
but see also \cite{Teyssier:2010ia} and references therein), although other
factors (orbit, gas fraction, morphology) are also important. Even though some
observations of IR luminous galaxies have failed to identify evidence for
mergers, it has been argued that these highly obscured objects could be the end
products of collisions between gas rich galaxies in a final phase of merger
evolution \citep[e.g.][]{Auriere:1996p1135}. Although this picture is well
supported in the local Universe, the physical origin of ULIRGs at high redshift
is less clear. For example, \cite{Sturm:2010p1055} identified two galaxies in
early \hersch observations that suggest that high redshift galaxies may be able
to achieve LIRG and even ULIRG luminosities without experiencing mergers. Our
model assumes that mergers can trigger bursts of star formation, but also
incorporates the high gas accretion rates at high redshift predicted by
cosmological simulations. It is therefore quite interesting to investigate what
our model predicts for the fraction of luminous IR galaxies that are the result
of mergers at high and low redshift.

To study the fraction of IR bright galaxies that are powered by mergers we
record the time since the last merger $T_{\mathrm{merger}}$ or major merger
$T_{\mathrm{majormerger}}$ for each galaxy in our simulation. ``Minor'' mergers
are defined as those with a mass ratio\footnote{Here, the masses used in the
ratio include stars, gas, and the dark matter within the optical radius of the
galaxy, as in S08} greater than 1:10, and ``major mergers'' as those with mass
ratio greater than 1:4. Our model predicts that about $84$ $(53)$ per cent of
high-redshift galaxies with $S_{160} > 10$ mJy have experienced a (major) merger
during their lifetime. Note, however, that the fraction drops to about $34$
percent if we concentrate on major mergers that have taken place less than $250$
Myr ago, which are the mergers that are most likely to be causally linked to the
star formation we are seeing. For the SPIRE 250 \mum band, the results are very
similar; about $86$ $(57)$ per cent of high-redshift galaxies with $S_{250} >
20$ mJy have experienced a (major) merger. If we again concentrate only on
recent mergers in which the major merging event took place less than $250$ Myr
ago, the merger fraction drops to $\sim 42$ per cent. Thus our model makes an
interesting prediction: roughly every second high-redshift ULIRG can achieve
ULIRG luminosities without a recent major merger. This prediction differs from
the findings of \cite{Gonzalez:2010p1071}, who used a semi-analytic galaxy
formation model \citep{Baugh:2005p1160} to study submillimetre galaxies (SMGs;
$S_{850} > 5$ mJy). \cite{Gonzalez:2010p1071} finds that the majority of SMGs
seen in current sub-mm surveys are starbursts triggered by major and minor
mergers involving gas-rich disk galaxies \citep[see also][]{Baugh:2005p1160}.

\begin{figure}
\includegraphics[width=\columnwidth]{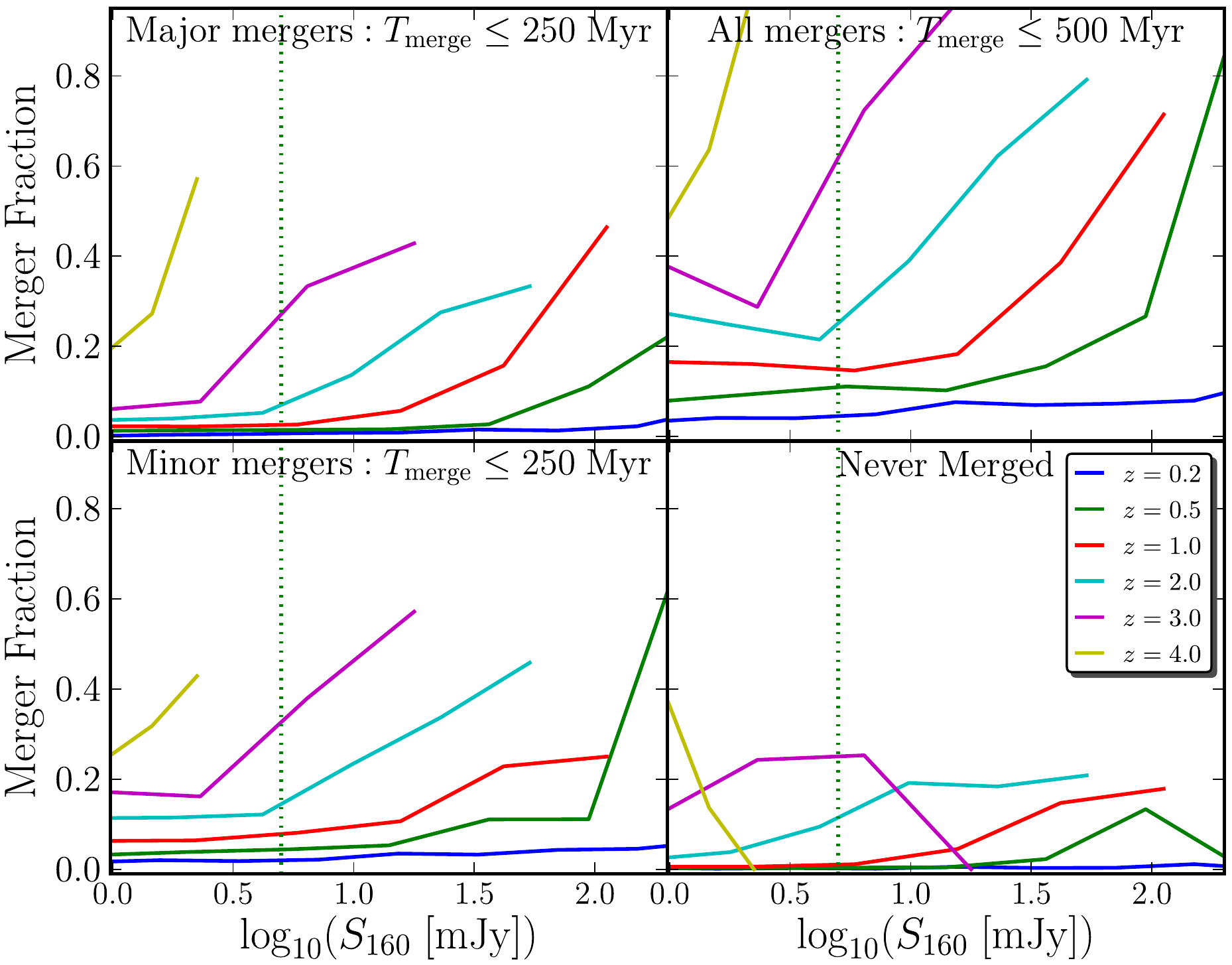}
\caption{Merger fraction as a function of PACS 160 \mum flux at different
redshifts. The upper-left plot shows the recent major mergers, while the
lower-left shows recent minor mergers (see text for definitions). The upper
right panel shows the merger fractions as a function of PACS 160 \mum flux for
all mergers within the past 500 Myr, and the lower right panel shows the
fraction of galaxies that have never merged.  The vertical dotted line shows a
representative flux limit for GOODS-\hersch (5 mJy).}
\label{fig:MergerFractionsSPIREflux}
\end{figure}

We present these results in more detail in Figure
\ref{fig:MergerFractionsSPIREflux}, which shows the evolution of merger
fractions as a function of PACS 160 \mum flux at different redshifts. The
upper-left plot shows that the merger fraction of recent (the latest merger took
place less than or equal to $250$ Myr ago) major mergers grows as a function of
PACS 160 \mum flux. This is true independent of studied redshift, although the
steepest rise is noted at $z = 1$ and $4$. The plot shows that the likelihood of
the most luminous ULIRGs being powered by a merger is higher than for those that
are less luminous. A similar trend can also be seen for minor mergers. This
shows that under the right conditions, minor mergers may also trigger
(U)LIRG-like activity. We note that Fig. \ref{fig:MergerFractionsSPIREflux} and
the related results are qualitatively similar if instead of the PACS 160 \mum
flux we use the SPIRE 250 \mum band.

We can use our models to attempt to identify observational quantities that could
provide interesting indicators of starburst and merger activity for observations
that are available or will become available soon.  We wish to exploit the
availability of extensive multi-wavelength data, from the rest UV to the FIR
with \herschn, in fields such as GOODS. We focus on the redshift range
\redshift\ for these experiments, and use the high redshift sample from our
lightcones. It is reasonable to expect that starbursts will have high values of
specific star formation rate (SSFR), $\dot{M}_{\star} \, M_{\star}^{-1}$. In a
relatively narrow redshift range, we can use the ratio of a FIR flux to a
Spitzer IRAC band as a proxy for SSFR.

\begin{figure}
\includegraphics[width=\columnwidth]{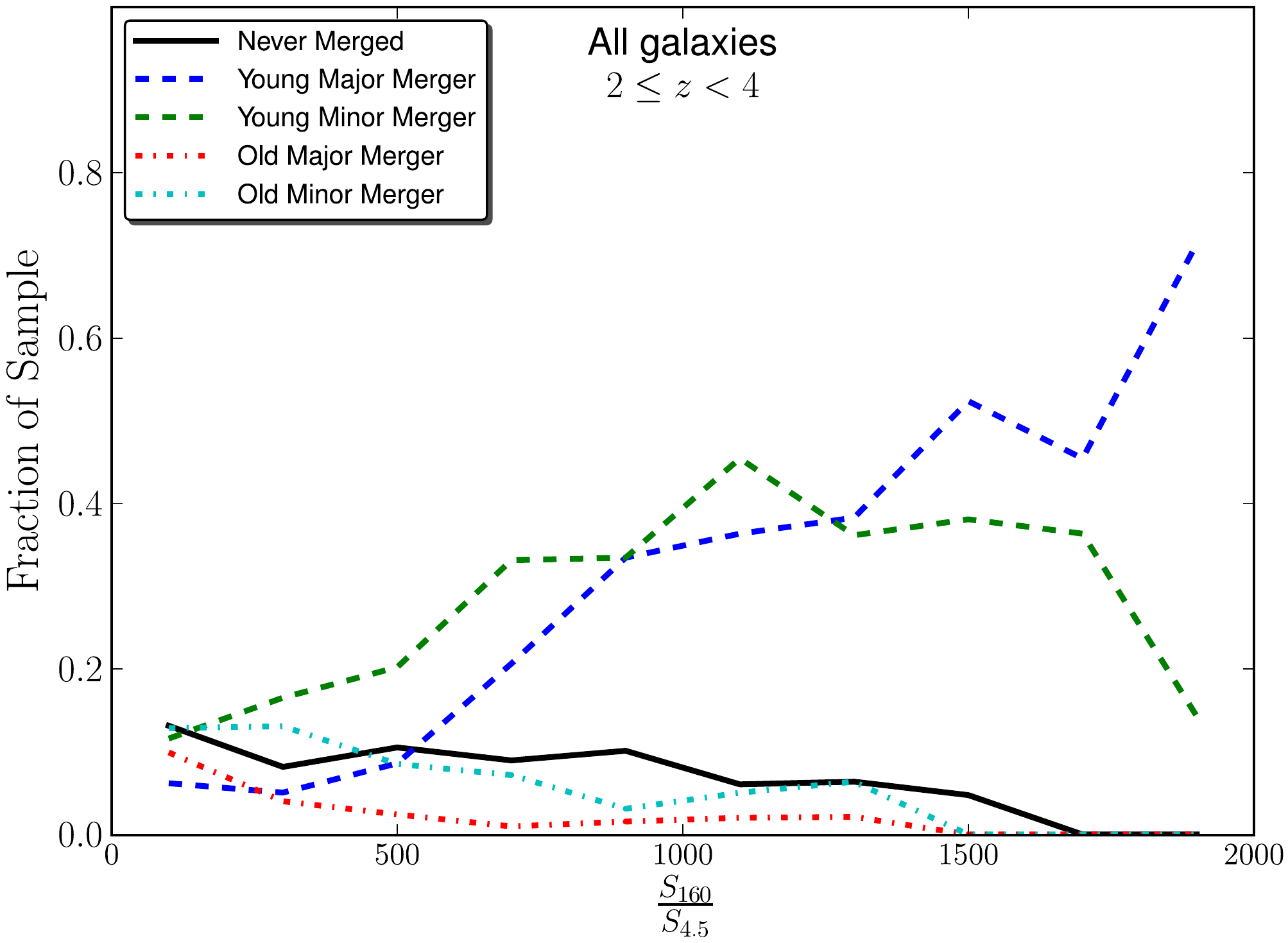}
\caption{The fraction of galaxies that have experienced a merger and that have
never merged with another galaxy as a function of far- to mid-IR flux ratio. The
flux ratio is the ratio between the PACS 160 and \spitz IRAC 4.5 \mum bands. The
merger events have been divided into recent or young ($\leq 250$ Myr) and less
recent or old ($250 < T_{\mathrm{merger}}< 500$ Myr) and to minor and major
incidents and are colour coded as given by the legend.}
\label{fig:SpitzerMergerFractions}
\end{figure}

Figure \ref{fig:SpitzerMergerFractions} shows the fraction of different types of
mergers and the fraction of galaxies which have never merged as a function of
far- to mid-IR flux ratio (colour). We choose to use the PACS 160 \mum band as
our far-IR band, because it is relatively close to the peak of the IR flux
distribution in the studied redshift range (\redshift), but at the same time it
suffers less from confusion than, e.g., the SPIRE 250 \mum band. The mid-IR band
chosen is the channel two of the \spitz IRAC at $4.5$ \mumn, which should
provide a proxy for the stellar mass of a galaxy. Fig.
\ref{fig:SpitzerMergerFractions} shows that our models predict that the
probability that a galaxy has experienced a recent merger is a strong function
of this far- to mid-IR flux ratio. For the most extreme values ($S_{160} \,
S_{4.5}^{-1} \gtrsim 1300$) more than half of the galaxies have experienced a
major merger within the last $250$ Myr.  We note that the results remain if we
instead use other bands such as $S_{250} \, S_{8}^{-1}$ as the flux ratio. It
will be interesting to test whether these IR colors are correlated with
signatures of recent merger activity in observed galaxies.

\begin{figure*}
\begin{minipage}{17cm}
\includegraphics[width=16cm]{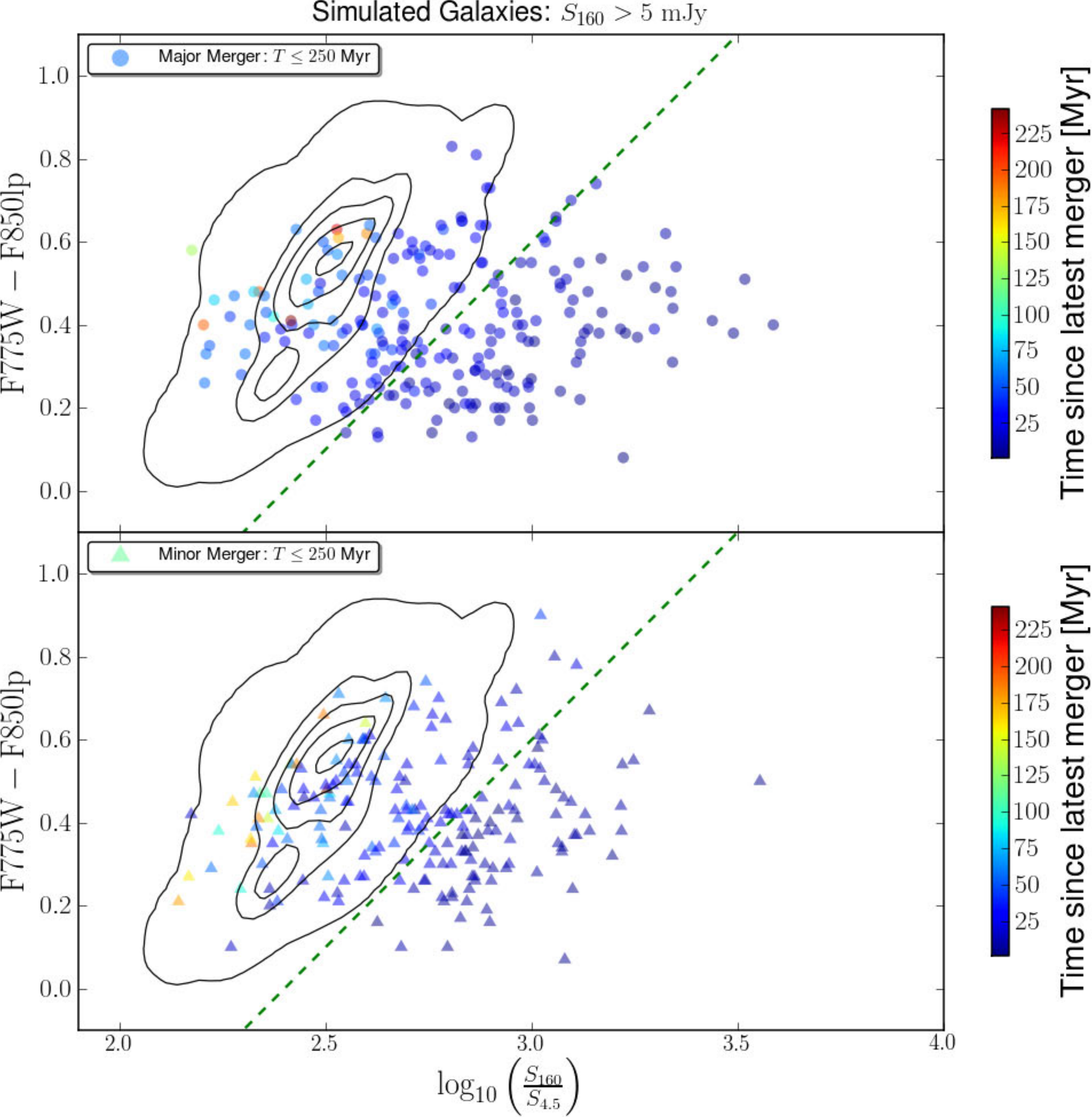}
\caption{Rest-frame UV-optical colour as a function of far- to mid-IR flux. The
flux ratio is the ratio between the PACS $160$ and \spitz IRAC $4.5$ \mum
bands. The model galaxies have been colour coded by the time since the last
merger in Myr. The green dotted line marks the region in which all simulated
galaxies are recent mergers.}
\label{fig:ColorMerger}
\end{minipage}
\end{figure*}

We can also explore whether the use of a two colour diagram can provide further
information. We expect that the rest-frame UV colours might correlate with the
age of the starburst, and thus with the time since the merger event. We consider
the filters of the Hubble Space Telescope's Advanced Camera for Surveys, because
such data are available for GOODS. In Figure \ref{fig:ColorMerger} we combine
the information on UV-optical colour and far- to mid-IR flux ratio in a
colour-colour plot. Again, we choose to use the PACS 160 \mum band and IRAC's
channel 2 ($4.5$ \mumn), because these channels allow the most robust comparison
with observations of galaxies in the GOODS field. However, the results would be
qualitatively similar with any IRAC channel or with the SPIRE 250 \mum band.
Fig. \ref{fig:ColorMerger} shows that in our models, the most recent mergers are
all located in the lower right region of each subplot, that is the region with
the most extreme far- to mid-IR flux ratio and the bluest UV-optical colour.

The black contours indicate where the galaxies which have never merged are
located and have been drawn using two dimensional Gaussian kernel density
estimator. Note, however, that this region is also populated by recent mergers
so we cannot conclude that the observed galaxies in this region have not
experienced a recent merger. However, it is interesting that there is a region
of the diagram (shown by a green dotted line) in which all galaxies have
experienced a recent merger. The functional form of the green line is as follows:
\begin{equation}
(F775W - F850lp)_{\mathrm{AB}} = A \log_{10}\frac{S_{160}}{S_{4.5}} + B	\quad ,
\end{equation}
where $A = 1$ and $B = -2.4$. Note, however, that this simple selection does not
seem to be able to separate recent major mergers from less extreme mergers. If
instead of the PACS we had used the SPIRE 250 \mum band, the above equation
would remain the same, but with $B = -2.7$.

It will be very interesting to test these predictions using surveys such as
GOODS-\herschn, where multi-wavelength photometry as well as high resolution
Hubble imaging are available. We plan to pursue this in a future work (Niemi et
al. in prep).

\section{Summary and Conclusions}\label{s:conclusions}

In this work we made use of the semi-analytic models developed by S11, in which
attenuation and re-emission of light by dust is modelled with a simple analytic
approach, to make predictions for the physical and observable properties of
galaxies that may be detected in deep fields by the \hersch Space Observatory.
We used the semi-analytic models to generate mock lightcones resembling the
GOODS survey but with 100 times the volume.

We first examined the differential number counts for galaxies in the PACS and
SPIRE bands. We found very good agreement with the observed number counts
derived from \hersch PACS data, for the overall counts and for galaxies binned
by redshift at $z<2$. At $z > 2$ our model underpredicts the number of more
luminous galaxies in the PACS 100 and 160 \mum bands. The agreement of our model
predictions with the observed counts deteriorates as one moves to longer
wavelengths, with increasingly poor agreement seen as one moves from PACS to
SPIRE 250 \mum and then to the longer SPIRE bands. At 250 \mum and longer
wavelengths, the models tend to underpredict the number counts of intermediate
flux galaxies ($10 \lesssim S_{250} \lesssim 100$ mJy), and overpredict the
counts of brighter galaxies ($S_{250} \gtrsim 100$ mJy). We presented
predictions using three different sets of empirical dust emission templates, and
found interesting differences. However, it does not appear that any single
simple modification to the dust templates could solve the discrepancies.

We discussed possible reasons for these discrepancies. These include
uncertainties due to cosmic variance, possible evolution or scatter in the dust
emission templates, a possible contribution to the IR light from obscured AGN,
and the possibility that many of the bright sources are lensed. It is also
possible that some of the physical processes in our model need revision, or that
the stellar IMF was different in early galaxies, as has been suggested by
\cite{Baugh:2005p1160}. However, our models produce reasonable agreement with
the luminosity functions at rest-UV through NIR wavelengths, and with the
observationally derived stellar mass functions and SFR functions at high
redshift \cite[see S11;][and references therein]{fontanot:09b}. Moreover, the
top-heavy IMF model of \citet{Baugh:2005p1160} that was developed to explain the
observed sub-mm population at 850 \mum does not appear to reproduce the SPIRE
number counts particularly well \citep{Clements:2010ce,Lacey:2010hp}, and does
not produce enough galaxies with large stellar masses at high redshift
\citep{Swinbank:2008eo}. We also discuss the impact of flux boosting and
blending on the observationally derived \hersch counts. This would explain why
the discrepancy becomes larger with increasing wavelength, and might also be a
larger effect for high redshift galaxies, which are compact and strongly
clustered. A quick visual inspection of HST images for high redshift galaxies
makes it clear that many, if not most, \hersch sources probably have
contributions from multiple galaxies. Moreover, according to the
simulations carried out by \cite{2011MNRAS.415.2336R}, more than half ($\sim 57$
per cent) of sources detected at $\geq 5\sigma$ at $500$ \mum show a flux
boosting by a factor $> 1.5$ and more than every fourth ($\sim 27$ per cent) by
a factor $> 2$. This is clearly a very serious issue when comparing theoretical
predictions with these observations. We intend to use our simulations to create
mock \hersch images, carry out the source detection procedure on these, and
study the predicted effects of blending on the measured fluxes and galaxy
counts. In addition, the \hersch team is working hard to better understand and
overcome this problem.

With the resulting caveat that one probably cannot interpret our
predicted \hersch fluxes as corresponding in a one-to-one fashion with
directly observed \hersch fluxes (as currently available), but rather as
intrinsic fluxes in the absence of blending, we used our model to make
predictions for physical properties of \hersch detected galaxies. We
showed the predicted relationships between PACS 160 and SPIRE 250 \mum
fluxes and stellar mass, halo mass, cold gas mass, SFR, and total IR
luminosity as various redshifts. Our models show fairly strong trends
between both $S_{160}$ and $S_{250}$ at a given redshift and all of
these quantities except halo mass, which is quite flat as a function of
both $S_{160}$ and $S_{250}$ at higher redshifts. This may have
interesting implications for the clustering properties of these sources
\citep[for early observational results in GOODS South, see
e.g.][]{2011arXiv1105.4093M}. Our models predict that the galaxies that
are likely to be detected in deep \hersch surveys such as GOODS-\hersch
($S_{160}$ or $S_{250} > 5$ mJy) at high redshift ($z>2$) have fairly
large stellar masses ($M_{\star} \sim 6 \times 10^{10} - 3 \times
10^{11}$\msun) and reside in fairly massive dark matter halos ($\gtrsim
10^{11}$\msun). The latter is in good agreement with the minimum mass
derived by \cite{Amblard:2011p1070} based on the observed clustering of
SPIRE sources.

The simulated IR luminous galaxies were also found to be gas rich with
cold gas masses ranging from $\sim 10^{9}$ up to $\sim 4 \times
10^{11}$\msun. The bulk of the high redshift IR bright galaxies were
found to contain $\sim 6 \times 10^{10}$\msun\ of cold gas. These
predictions will soon be able to be tested with radio and mm
observations such as those that will be obtainable with ALMA.

Our predicted stellar, cold gas, and halo masses at a given 250 \mum
flux are significantly higher than the corresponding predictions from
the model of \citet{Lacey:2010hp}, which assumes a top-heavy IMF in
starbursts. We show a quantitative comparison between our predicted SFR
vs. $S_{250}$ relation and that of \citet{Lacey:2010hp}, finding that
although their predictions are similar to ours for low redshift
galaxies, at higher redshift ($z\gtrsim 0.5$), their predicted SFR are
lower than ours (at a given FIR flux) by about a factor of two.

We investigated the radial sizes of high redshift (\redshift) late-type galaxies
selected in the SPIRE 250 \mum band and compared our predictions with the
observations recently presented by \citet{Cava:2010p1050}. We found that the
SPIRE detectable galaxies are on average significantly larger than IR faint
galaxies with comparable stellar masses. The mean disk size of IR luminous
late-type galaxies was found to be in good agreement with the observations,
although the statistics are extremely limited. We speculate that this is because
larger galaxies are more gas rich and therefore more likely to be IR bright.

Finally, we investigated our model predictions for the importance of
merger-driven starbursts in producing IR luminous galaxies at different
redshifts. Although there is clear evidence for a connection between merger
activity and LIRGs and ULIRGs in the local Universe, the situation at high
redshift is less clear. However, it has been suggested that an increasing
fraction of LIRGs and ULIRGs at high redshift might not have a merger origin. We
present quantitative predictions for the fraction of galaxies that have
experienced a major or minor merger within a given timescale as a function of
redshift and 160 \mum flux. We find a fairly strong trend between the 160 \mum
flux and the probability that a galaxy has had a recent merger, indicating that
brighter galaxies are more likely to be merger driven. However, we find the
interesting result that in our models, a significant fraction (half or more) of
IR-luminous galaxies at high redshift ($z>2$) have not experienced a recent
merger. This implies that the high gas accretion rates and efficient feeding via
cold flows predicted by cosmological simulations at high redshift can fuel a
significant fraction of the galaxies detected by \herschn. This appears
consistent with preliminary observational results
\citep[e.g.][]{Sturm:2010p1055}, but will be interesting to study and quantify
in more detail in the near future.

In order to pursue this question further, we used our model to try to identify
some observational photometric signatures associated with galaxies that have
experienced recent mergers. We showed that there is a strong correlation between
the mid-to-IR colour $S_{160}/S_{4.5}$ (PACS 160 \mum to \spitz IRAC channel 2
ratio) and the probability that a galaxy has had a recent merger. We also
investigated a two color diagnostic using rest-frame UV colour and the 160 \mum
to IRAC 4.5 \mum ratio. We found that recent mergers occupy a characteristic
region of this diagram, suggesting another interesting observational tool to
identify probable recent mergers. In a work in progress, we are making use of
HST imaging in the GOODS-\hersch survey to study the morphologies of galaxies
selected using these multi-wavelength diagnostics, in order to test our model
predictions and the importance of merger-induced star formation at high redshift
(Niemi et al., in prep).

\section*{Acknowledgements}

We thank Matthieu B\'ethermin, David Elbaz, Mark Dickinson, Jeyhan Kartaltepe,
Carolin Villforth, Denis Burgarella, and Alexandra Pope for useful comments and
stimulating discussions. We thank the anonymous referee for comments that helped
us to improve the paper.

\bibliographystyle{mn2e} 
\bibliography{fixed}

\bsp

\label{lastpage}

\end{document}